\documentclass[superscriptaddress,prb,amsmath,amssymb,notitlepage,twocolumn]{revtex4-1}
\usepackage[pdftex]{graphicx}
\usepackage{subfigure}
\usepackage{dcolumn}
\usepackage{bm}
\usepackage[usenames,dvipsnames,svgnames]{xcolor}
\usepackage[colorlinks]{hyperref}
\hypersetup{
    linkcolor=BrickRed,
    urlcolor=blue,
    citecolor=MidnightBlue,
}

\newcommand{\ket}[1]{\vert #1 \rangle}

\renewcommand{\vec}[1]{{\mathbf #1}}

\newcommand{\ee}{\'{e}e}

\newcommand{\UIUCECE}[0]{\affiliation{Department of Electrical and Computer Engineering, University of Illinois at Urbana-Champaign, Urbana, IL 61801, USA}}

\newcommand{\UIUCMNTL}[0]{\affiliation{Micro and Nanotechnology Laboratory, University of Illinois, 208 N. Wright Street, Urbana IL 61801, USA}}
\newcommand{\UIUCMATSE}[0]{\affiliation{Department of Materials Science and Engineering, University of Illinois at Urbana-Champaign, Urbana, IL 61801, USA}}
\newcommand{\UIUCFS}[0]{\affiliation{Frederick Seitz Materials Research Laboratory, University of Illinois at Urbana-Champaign, Urbana, IL 61801, USA}}
\newcommand{\UIUCNCSA}[0]{\affiliation{National Center for Supercomputing Applications, University of Illinois at Urbana-Champaign, Urbana, IL 61801, USA}} 

\bibpunct{(}{)}{;}{s}{,}{,}

\begin{document}

\title{Voltage induced switching of an antiferromagnetically ordered topological Dirac semimetal}

\author{Youngseok Kim}\UIUCECE\UIUCMNTL
\author{Kisung Kang}\UIUCMATSE
\author{Andr\'{e} Schleife}\UIUCMATSE\UIUCFS\UIUCNCSA
 \author{Matthew J. Gilbert}\UIUCECE\UIUCMNTL

\date{\today}

\begin{abstract}
An antiferromagnetic semimetal has been recently identified as a new member of topological semimetals that may host three-dimensional symmetry-protected Dirac fermions. 
A reorientation of the N\ee{l} vector may break the underlying symmetry and open a gap in the quasi-particle spectrum, inducing the (semi)metal-insulator transition. Here, we predict that such transition may be controlled by manipulating the chemical potential location of the material. We perform both analytical and numerical analysis on the thermodynamic potential of the model Hamiltonian and find that the gapped spectrum is preferred when the chemical potential is located at the Dirac point. As the chemical potential deviates from the Dirac point, the system shows a possible transition from the gapped to the gapless phase and switches the corresponding N\ee{l} vector configuration. We perform density functional theory calculations to verify our analysis using a realistic material and discuss a two terminal transport measurement as a possible route to identify the voltage induced switching of the N\ee{l} vector.  
\end{abstract}


\maketitle

\section{Introduction} \label{sec:intro}

Graphene has long been studied as a condensed matter testbed for the relativistic Dirac equation\cite{Geim2009,Novoselov2012}. 
The Dirac points, where the valence and the conduction band touch, are protected from opening a gap by the combination of time-reversal ($\mathcal{T}$) and inversion ($\mathcal{P}$) symmetry\cite{Geim2009}. 
The Dirac points are robust to symmetry preserving perturbations providing a reliable platform to study many intriguing physical phenomena including an extremely high carrier mobility\cite{Novoselov2004, Novoselov2005}, Klein tunneling\cite{Katsnelson2006,Young2009}, and anomalies in quantum Hall effect\cite{Zhang2005,Novoselov2006,Novoselov2007}. 
Recent progress in topological semimetals has extended the realization of relativistic particles in a condensed matter system from two-dimensional to three-dimensional solids\cite{Armitage2017}. 

A linear crossing of the band inside the bulk has been predicted and discovered in materials with a broken\cite{Huang2015,Weng2015,Liang2017} $\mathcal{P}$ or broken\cite{Wan2011, Burkov2011} $\mathcal{T}$. 
The low-energy excitation of such materials is described by Weyl fermions, whose Hamiltonian is $H(\vec{k})=\sum_{i,j=x,y,z} v_{ij} k_i\sigma_j$, where $\sigma_{i=x,y,z}$ are the Pauli matrices and $v_{ij}$ is the Fermi velocity assuming $\det[v_{ij}]\neq0$. Weyl fermions cannot be gapped out by perturbations as all the Pauli matrices are used in its construction, and these band crossing points, Weyl nodes, are characterized by the Chern number obtained from\cite{Rappe2012} $\text{sgn}(\det[v_{ij}])=\pm1$. Nevertheless, when two Weyl fermions with opposite Chern numbers are brought together at the same point in momentum space, such four-fold degeneracy is not protected and may be gapped out, annihilating the Weyl nodes. However, a particular space group symmetry along with the $\mathcal{P}$ and $\mathcal{T}$ symmetry may prohibit such annihilation and preserve the four-fold degeneracy, forming a 3D Dirac fermion\cite{Rappe2012, Young2015}. Although breaking any of the above symmetries typically lifts the four-fold degeneracy\cite{Armitage2017}, breaking both $\mathcal{P}$ and $\mathcal{T}$ but preserving the combination of $\mathcal{PT}$ along with an additional space group symmetry may still protect the Dirac fermion\cite{Tang2016}. 
An antiferromagetic semimetal (AFS) is an example where both $\mathcal{P}$ and $\mathcal{T}$ are broken but $\mathcal{PT}$ may be preserved with an extra non-symmorphic crystal symmetry. Recent studies have shed light on AFS as a new class of materials that may host symmetry protected Dirac fermions\cite{Tang2016, Wang2017_1, Jungwirth2017}. Interestingly, the reorientation of the antiferromagnetic (AF) order may break the underlying crystal symmetry and gap out Dirac fermions in the presence of spin-orbit coupling. Then the system may undergo a (semi)metal-insulator transition (MIT) resulting in a change in the magnetoresistance of the material\cite{Jungwirth2017}. Therefore, the electronic transport response of AFS may be modulated by manipulating the spin orientation, thereby providing a potential novel platform for spintronic devices and applications\cite{Jungwirth2017}.

A straightforward method to control the magnetism is to couple the net magnetic order of the material with the external magnetic field. However, the lack of net magnetization in antiferromagnets makes it challenging to find a suitable knob to control antiferromagnetic order. In this regard, spin-orbit coupling provides a promising path to control the spin degree of freedom via charge current flow. Such spin transfer torque is referred to as \emph{spin-orbit torque} (SOT)\cite{Gambardella2011, Jungwirth2016} and induces non-equilibrium spin polarization even in the absence of a net magnetization. Recently, it has been theoretically proposed\cite{Jungwirth2014, Jungwirth2017} and experimentally demonstrated\cite{Wadley2016} that current induced SOT provides a novel way to manipulate AF order. Although SOT similarly provides a mean to manipulate AF order of the AFS, it is unclear whether a sufficient SOT is induced when AFS is initially in a gapped phase due to the lack of the current flow. In addition, biaxial anisotropy energy is required to stabilize two distinct spin configurations which correspond to the metallic and insulating phase of the AFS. To tackle this challenge, we propose an alternative route to induce MIT by utilizing a gate controlled AF order. The mechanism we propose exploits the fact that the gapless and gapped phase of the AFS exhibit different total energy depending on the location of the chemical potential, therefore, we may manipulate a preferred phase and its corresponding AF order by gating the material.
Such voltage induced magnetic order control has been proposed in a ferromagnetic material proximity coupled to the graphene\cite{Kiwook2008} and topological insulator surface states\cite{Kiwook2012,Kiwook2017}, and has been proposed conceptually in antiferromagnetic material\cite{Libor2018}.

The remainder of the paper is organized as follows. In Section~\ref{sec:DSM}, we first introduce a background on symmetry protected Dirac fermions in AFS. In Section~\ref{sec:TB}, we outline the tight-binding model Hamiltonian that we will use for numerical calculation of the AFS properties. In Section~\ref{sec:Free}, we describe our analysis on the free energy of different AF order configurations and show that the MIT occurs by manipulating the chemical potential of the system. In Section~\ref{sec:current}, we additionally propose a two-terminal experiment on the AFS proximity coupled with a ferromagnetic insulator which exhibits a distinct signature of the voltage induced AF order switching. In Section~\ref{sec:Con}, we summarize our results.

\section{Symmetry protected Dirac semimetal in Antiferromagnetic material} \label{sec:DSM}

Orthorhombic CuMnAs and CuMnP have been investigated as candidate materials that may host 3D Dirac fermions near the Fermi level\cite{Tang2016}. The eigenvalue spectrum indeed shows multiple Dirac fermions along the edge of the Brillouin zone (BZ). While the Dirac fermions are gapped in the presence of spin-orbit coupling, the Dirac points located at specific BZ edge are protected. Understanding the physics of such protected Dirac fermions is a crucial step toward realizing MIT in AFS. 

To this end, two key symmetries are considered here: the $\mathcal{PT}$ symmetry and non-symmorphic symmetry. The $\mathcal{PT}$ symmetry satisfies $(\mathcal{PT})^2=-1$ which guarantees at least one degenerate states at high-symmetry momenta by Kramer's theorem. Due to the fact that the eigenstate and its $\mathcal{PT}$ symmetric partner shares the same momentum, the bands are two-fold degenerate in the entire BZ. These doubly degenerate states are a prerequisite for the Dirac semimetal, as the Dirac point is comprised of a combination of two doubly degenerate Weyl points. However, in general, such crossing results in a gap due to the level repulsion unless we introduce an additional symmetry to \emph{stick}\cite{Wang2017_2} the bands together at the band crossing point. 

To expand our discussion on the role of this additional symmetry, we assume an arbitrary Hamiltonian that respects $\mathcal{PT}$ and an additional non-symmorphic symmetry. The $\mathcal{PT}$ symmetry satisfies $(\mathcal{PT})^2=-1$ and has eigenvalues of $\lambda_{\pm}=\pm i$. 
Unlike the $\mathcal{PT}$ symmetry, the non-symmorphic symmetry produces an extra phase factor in its eigenvalue due to the partial translation of the lattice vector. We may see this additional phase factor by defining the glide mirror symmetry as $\mathcal{G}_x=\{M_x|\frac{1}{2}00\}$ which is a combination of a mirror symmetry followed by a translation of the real space coordinate as $(x,y,z)\rightarrow(x+\frac{1}{2},y,z)$. 
In general, the glide mirror symmetry $\mathcal{G}_x=\{M_x|\vec{a}\}$ produces\cite{Wang2017_2} $\mathcal{G}_x^2=-T(2\vec{a}_\parallel)=-e^{-\vec{k}\cdot 2\vec{a}_\parallel}$, where $\vec{a}$ is a fractional primitive translation lattice vector, $\vec{a}_\parallel$ is a projection of $\vec{a}$ onto the mirror plane, $T(\vec{x})$ is a translation operator, and a minus sign appears due to the $2\pi$ rotation of the spin. As a result, the eigenvalue for $\mathcal{G}_x$ is $g_\pm=\pm ie^{-\vec{k}\cdot \vec{a}_\parallel}$ with an additional phase factor. In our example, $\vec{a}=(\frac{1}{2}00)$ and $\vec{a}_\parallel=(000)$, thus the eigenvalue for $\mathcal{G}_x$ becomes $g_{\pm}=\pm i$.

Having both $\mathcal{PT}$ and $\mathcal{G}_x$ operators defined, we wish to examine the eigenvalue of the two-fold degenerate states. If two-fold degenerate states share the same eigenvalue, we may observe a protected Dirac point. To explain this, we first assume the Bloch state $\ket{\psi_\mathbf{k}^+}$ which is an eigenstate of $\mathcal{G}_x$ with eigenvalue $g_+=i$. To be more explicit, we wish to evaluate the eigenvalue of its degenerate partner, $\mathcal{PT}\ket{\psi_\mathbf{k}^+}$.
We first note that the real-space coordinate is transformed upon the operation of the following sequence of the operators:
\begin{equation}
\begin{split}
&(x,y,z)
\xrightarrow{\mathcal{G}_x}
(-x+\frac{1}{2},y,z)
\xrightarrow{\mathcal{PT}}
(x-\frac{1}{2},-y,-z), \\
&(x,y,z)
\xrightarrow{\mathcal{PT}}
(-x,-y,-z)
\xrightarrow{\mathcal{G}_x}
(x+\frac{1}{2},-y,-z). \\
\end{split}
\end{equation}
In other words, $\mathcal{G}_x\mathcal{PT}=T(1,0,0)\mathcal{PTG}_x=e^{-ik_x}\mathcal{PTG}_x$, where $T$ is a translation operator. 
Then, the degenerate partner, $\mathcal{PT}\ket{\psi_\mathbf{k}^+}$, satisfies\footnote{Equation~(\ref{eq:Gx}) is obtained by following the same procedure outlined in Eq. (16) of the supplementary material in \citet{Tang2016}.}
\begin{equation} \label{eq:Gx}
\begin{split}
\mathcal{G}_x\{\mathcal{PT}\ket{\psi_\mathbf{k}^+} \}
=&e^{-ik_x}g_- \{\mathcal{PT}\ket{\psi_\mathbf{k}^+}\}, \\
\end{split}
\end{equation}
where the extra phase factor is from $\mathcal{G}_x\mathcal{PT}=e^{-ik_x}\mathcal{PTG}_x$.
At $\vec{k}=(0,k_y,k_z)$, the degenerate states have different eigenvalues for $\mathcal{G}_x$ ($g_+$ for $\ket{\psi_\mathbf{k}^+}$ and $g_-$ for $\mathcal{PT}\ket{\psi_\mathbf{k}^+}$). In this case, a crossing with any other eigenstate either with $g_+$ and $g_-$ causes a level repulsion and, thus the Dirac points are not protected\cite{Tang2016}. At $\vec{k}=(\pi,k_y,k_z)$, the degenerate states have the same eigenvalue for $\mathcal{G}_x$ ($g_+$ for both $\ket{\psi_\mathbf{k}^+}$ and $\mathcal{PT}\ket{\psi_\mathbf{k}^+}$). As a result, there is no level repulsion if a crossing occurs with another degenerate state having an eigenvalue of $g_-$.  Therefore, the Dirac point is protected at the edge of the BZ, or specifically at $k_x=\pi$ in this example. 

\section{Model Hamiltonian} \label{sec:TB}

\subsection{Tight-binding model in real-space} \label{sec:TBR}
\begin{figure}[t!] 
  \centering
   \includegraphics[width=0.5\textwidth]{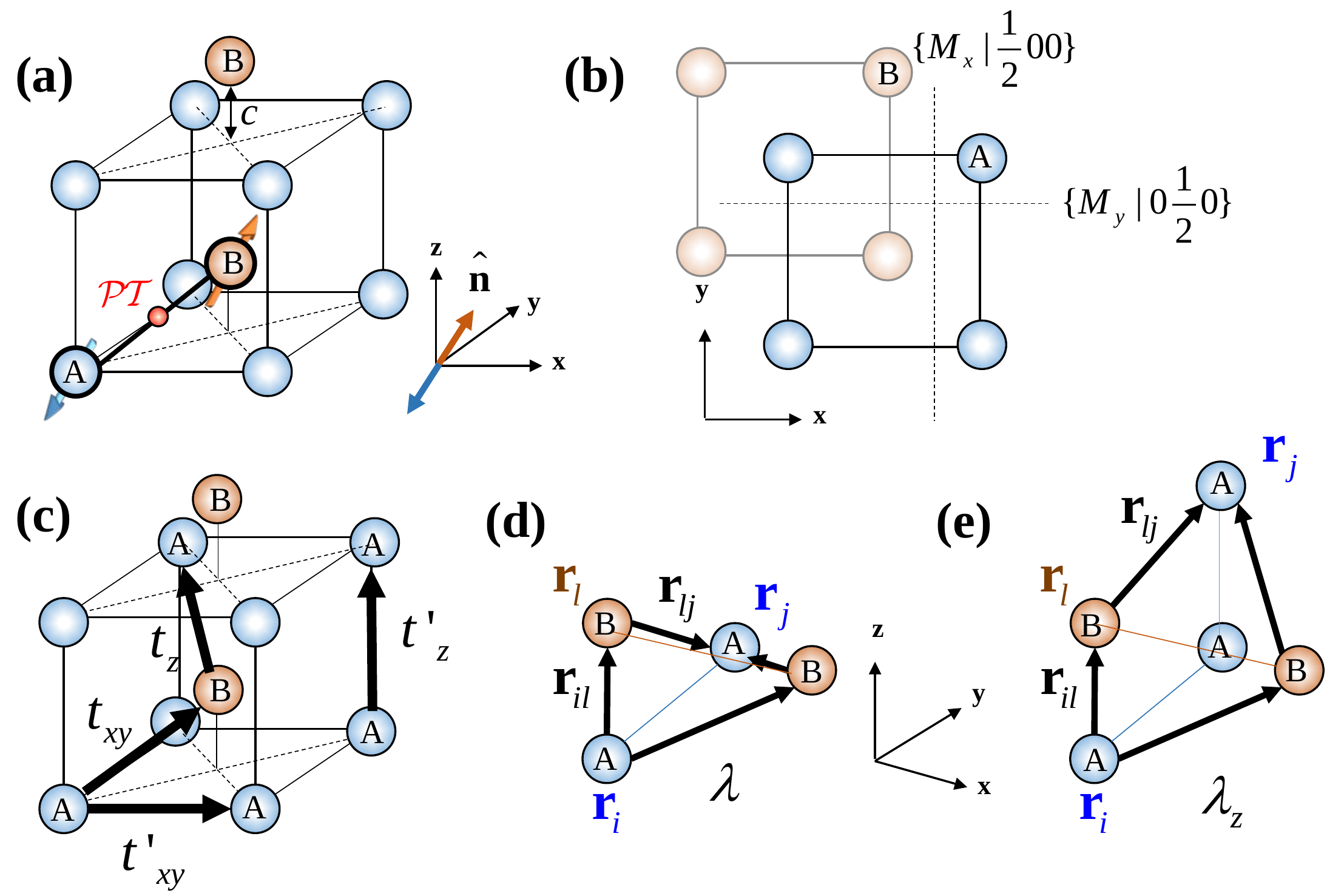}
  \caption{(a) A lattice structure consists of two sublattice atoms. Two sublattice atoms are indicated as A and B having an opposite spin configuration along the N\'{e}el vector $\hat{\bm{n}}$ (AF order) enforced by the exchange coupling. (b) The lattice structure has non-symmorphic symmetry $\mathcal{G}_x$ for $\hat{\bm{n}}||[100]$ and $\mathcal{G}_y$ for $\hat{\bm{n}}||[010]$. (c) The tight-binding model intra-layer and inter-layer hopping parameters. (d) The intra-layer momentum dependent spin-orbit coupling parameter, $\lambda$. (e) The inter-layer momentum dependent spin-orbit coupling parameter, $\lambda_z$. }\label{fig:lattice}
\end{figure}

To have insight on the low-energy Hamiltonian description, we now construct an effective tight-binding model for AFS system. The model we consider here has a tetragonal primitive structure whose lattice vector in 3D is defined as\cite{Wang2017_2, Jungwirth2017} $\vec{a}_1=(100)$, $\vec{a}_2=(010)$, and $\vec{a}_3=(001)$. Fig.~\ref{fig:lattice}(a) shows the lattice structure which has two sublattice atoms labeled as A and B, forming a bipartite square lattice in $\hat{x}-\hat{y}$ plane. In addition, Fig.~\ref{fig:lattice}(a) shows that the A and B atoms are buckled in opposite $\hat{z}$ direction with a constant displacement $\pm \frac{c}{2}\hat{z}$ ($0<c<1$). The AF order exists by enforcing the opposite spin direction for A and B sublattices. Such alternating spin order in bipartite lattice forms a N\ee{l} order and is aligned with a unit vector defined as the N\'{e}el vector $\hat{\vec{n}}$. 
In this lattice structure, $\mathcal{T}$ is broken by the magnetic order and $\mathcal{P}$ is broken due to the bipartite square lattice with a staggered spin order. However, the lattice preserves $\mathcal{PT}$ with an inversion center indicated as a red dot in Fig.~\ref{fig:lattice}(a), and all bands have two-fold degeneracy in the entire Brillouin zone. In addition, the lattice has additional non-symmorphic symmetry that consists of a mirror reflection symmetry followed by a half-lattice vector translation. Fig.~\ref{fig:lattice}(b) shows the corresponding non-symmorphic symmetries $\mathcal{G}_x=\{M_x|\frac{1}{2}00\}$ and $\mathcal{G}_y=\{M_y|0\frac{1}{2}0\}$ with their mirror plane indicated with a dashed line. 

The tight-binding Hamiltonian is\cite{Wang2017_2}
\begin{equation} \label{eq:tb1}
\begin{split}
H=&\sum_{\langle i j\rangle}t_{ij}c_i^\dagger c_j
+\sum_{\langle\langle i j \rangle\rangle,\langle l \rangle} c_i^\dagger i\lambda_{ij}(\hat{\mathbf{d}}_{il}\times \hat{\mathbf{d}}_{lj})\cdot \bm{\sigma} c_j \\
&+\Delta\sum_i \xi_i c_i^\dagger \bm{\sigma}\cdot \hat{\bm{n}},
\end{split}
\end{equation}
where $c_i=(c_{i\uparrow},c_{i\downarrow})^T$ is the electron annihilation operator at site $i$ located at $\bm{r}_i$, $\langle ij \rangle$ indicates the hopping range from nearest-neighbor to next-nearest neighbor sites with a corresponding hopping parameter $t_{ij}$. Fig.~\ref{fig:lattice}(c) shows the intra-plane ($t_{xy}$) and inter-plane ($t_z$) nearest-neighbor hopping between A and B sublattices. Fig.~\ref{fig:lattice}(c) also shows the intra-plane ($t_{xy}'$) and inter-plane ($t_z'$) next nearest-neighbor hopping between the same sublattices. 
The second term in Eq.~(\ref{eq:tb1}) is the next-nearest neighbor, or Kane-Mele $k$-dependent spin-orbit coupling (SOC)\cite{Kane2005, Jairo2017}. In Eq.~(\ref{eq:tb1}), $\bm{\sigma}=(\sigma_x,\sigma_y,\sigma_z)$ are the Pauli matrices for spin degree of freedom, $\langle\langle ij \rangle\rangle$ are the indices for the next-nearest neighbor atoms, $\hat{\bm{d}}_{il}=(\mathbf{r}_i-\mathbf{r}_l)/\vert\mathbf{r}_i-\mathbf{r}_l\vert$ is a unit vector that connects the target atoms at $i$ (or $j$) with the intermediate atom at $l$, located at the same distance from the site $i$ and $j$. Fig.~\ref{fig:lattice}(d) and (e) depicts the intra- and inter-layer SOC with SOC strength of $\lambda$ and $\lambda_z$, respectively. 
The last term in Eq.~(\ref{eq:tb1}) is the staggered Zeeman term with $\xi_i=-1$ for the sublattice A and $\xi_i=1$ for the sublattice B. In particular, we assume checkerboard N\ee{l} order and opposite direction of the spin at A and B sites are aligned with the N\'{e}el vector $\hat{\bm{n}}$.

\begin{figure*}
  \centering
   \includegraphics[width=1.0\textwidth]{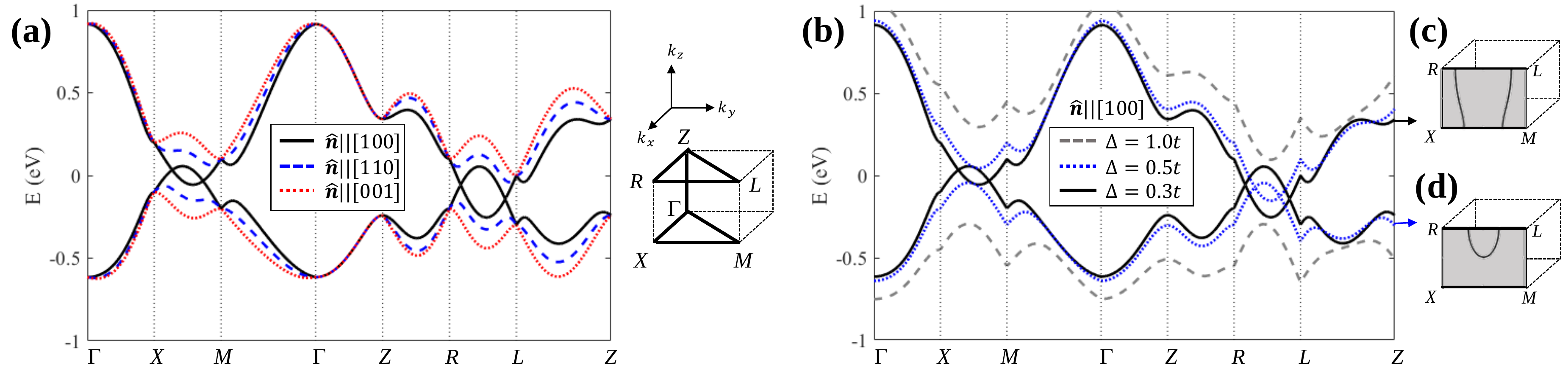}
  \caption{(a) The eigenvalue spectrum of the Hamiltonian in Eq.~(\ref{eq:tb2}). We use $t_{xy}=t=0.5$ eV, $t_{z}=0.5t$, $t'_{xy}=0.1t$, $t'_z=0.1t$, $\lambda=0.5t$, $\lambda_z=0.1t$, and $\Delta=0.3t$. We plot the spectrum for three different N\ee{l} vector alignments $\hat{\bm{n}}||[100]$ (black solid line), $\hat{\bm{n}}||[110]$ (blue dashed line), and $\hat{\bm{n}}||[001]$ (red dotted line). (b) The eigenvalue spectrum of the Hamiltonian with the same parameter choices as in (a), but with different exchange coupling energy $\Delta$. The black solid line has $\Delta=0.3t<\lambda-\lambda_z$, which has two DNLs at $k_x=\pi$ plane as it is shown in right upper side inset. The blue dotted line has $\lambda-\lambda_z<\Delta=0.3t<\lambda+\lambda_z$, having one DNL at $k_x=\pi$ plane. As a result, the $X-M$ line shows a gapped spectrum. The gray dashed line has $\Delta>\lambda+\lambda_z$ and the spectrum is fully gapped by the exchange energy although the gliding symmetry $\mathcal{G}_x$ is preserved.}\label{fig:spectrum}
\end{figure*} 

\subsection{Tight-binding model in momentum-space} \label{sec:TBK}
We Fourier transform the Hamiltonian in Eq.~(\ref{eq:tb1}) into momentum space to obtain\cite{Wang2017_2}
\begin{equation} \label{eq:tb2}
\begin{split}
H(\vec{k})=&[t_{xy}\tau_1+t_z(\tau_1\cos k_z+\tau_2\sin k_z)]\cos\frac{k_x}{2}\cos\frac{k_y}{2}\\
&+t'_{xy}(\cos k_x + \cos k_y) + t'_z\cos k_z + \Delta \tau_3 \mathbf{\sigma}\cdot \hat{\mathbf{n}} \\
&+(\lambda-\lambda_z\cos k_z)\tau_3(\sigma_2\sin k_x -\sigma_1 \sin k_y),
\end{split}
\end{equation}
where $\tau_{i=x,y,z}$ and $\sigma_{i=x,y,z}$ are the Pauli matrices for sublattice and spin degree of freedom, respectively. Following the parameters defined in Eq.~(\ref{eq:tb1}), $t_{xy}$ and $t'_{xy}$ are the intra-layer nearest- and next nearest-neighbor hopping parameter, $t_z$ and $t'_z$ are the inter-layer nearest- and next nearest-neighbor hopping parameters, $\lambda$ and $\lambda_z$ are the intra- and inter-layer SOC strength, respectively, and $\Delta$ is the staggered Zeeman energy of the AF order. We assume a layered structure, thus the hopping in the inter-layer direction is weaker than the intra-layer direction ($t_{xy}\geq t_z$, $t'_{xy} \geq t'_z$, and $\lambda \geq \lambda_z$). 


When the N\'{e}el vector is aligned with $\hat{x}$ direction ($\hat{\bm{n}}||[100]$), the Dirac points are protected by the gliding mirror reflection symmetry $\mathcal{G}_x$ according to the discussions in Section~\ref{sec:DSM}. Given the N\'{e}el vector configuration ($\hat{\bm{n}}||[100]$), we now wish to examine the relationship between SOC and the Zeeman energy to determine the existance and the location of the protected Dirac points at $k_x=\pi$.
To obtain a better understanding of the phase diagram, we begin by examining the parameter range of the $\lambda$ and $\Delta$, which allows the existence of the protected Dirac points in our model Hamiltonian. By setting $k_x=\pi$ and $\hat{\bm{n}}||[100]$ in Eq.~(\ref{eq:tb2}), the two-fold degenerate energy spectrum is obtained as
\begin{equation} \label{eq:E2}
\begin{split}
E_{\pm}(\pi,k_y,k_z)=&
t'_{xy}(\cos k_y-1) + t'_z\cos k_z \\
&\pm
(\Delta - (\lambda-\lambda_z\cos k_z)\sin k_y).
\end{split}
\end{equation}
Eq.~(\ref{eq:E2}) needs to satisfy $\Delta - (\lambda-\lambda_z\cos k_z)\sin k_y=0$ at the Dirac point, as a four-fold degeneracy is expected. Equivalently, the Dirac points or Dirac nodal line (DNL) may be found at momentum $\vec{k}=(\pi,k_y,k_z)$ which satisfies $\sin k_y=\Delta/(\lambda-\lambda_z\cos k_z)$. 
In case of $|\Delta|<\lambda-\lambda_z$, two $k_{y1}$ and $k_{y2}$ exist for a given $k_{z1}$ thus we find two DNLs in $k_x=\pi$ plane. If $\lambda-\lambda_z<|\Delta|<\lambda+\lambda_z$, one finds $k_{y1}$ and $k_{y2}$ within a specific range of $k_{z1}$ that satisfies $|\Delta|<\lambda-\lambda_z\cos k_{z1}$ and, as a result, we find one DNL in $k_x=\pi$ plane. Lastly, the spectrum is fully gapped when $|\Delta|>\lambda+\lambda_z$. 
In summary, the conditions between the SOC strength $\lambda$, and the Zeeman energy $\Delta$, are as follows: 
\begin{equation} \label{eq:DeltaCond}
\begin{split}
\vert\Delta\vert < \lambda-\lambda_z 
&\rightarrow\text{two DNLs} \\
\lambda-\lambda_z < \vert\Delta\vert < \lambda+\lambda_z 
&\rightarrow\text{one DNL} \\
\vert\Delta\vert > \lambda+\lambda_z
&\rightarrow\text{fully gapped}. \\
\end{split}
\end{equation}

As we now know the conditions for finding the Dirac points, let us obtain a low-energy effective model near the Dirac point to explicitly show that the Dirac point is protected by the non-symmorphic symmetry\cite{Wang2017_2} $\mathcal{G}_x$. Assuming that a Dirac point is located at $\vec{k}_1=(\pi,k_{y1}, k_{z1})$, the low-energy effective model is\cite{Wang2017_2}
\begin{equation} \label{eq:Hlow}
H(\vec{k}_1+\vec{q})=
(v_{x1}\tau_1+v_{x2}\tau_2+v_{x3}\tau_3\sigma_2)q_x
+(v_y q_y+v_z q_z)\tau_3\sigma_1,
\end{equation}
where $\vec{q}=(q_x,q_y,q_z)$ is a small deviation from $\vec{k}_1$, and $v_{x1},\;v_{x2},\;v_{x3},\;v_{y},\;v_z$ are obtained from Eq.~(\ref{eq:tb2}) and details of the calculation is given in Appendix~\ref{sec:Heff}. In Eq.~(\ref{eq:Hlow}), we clearly observe a linear dispersion near $\vec{q}=0$. In the presence of $\mathcal{PT}$ symmetry, $\mathcal{PT}=i\tau_1\sigma_2\mathcal{K}$, the allowed Gamma matrices that satisfy $[\mathcal{PT},\Gamma]=0$ are $\bm{\Gamma}=(\tau_1,\tau_2,\tau_3\sigma_1,\tau_3\sigma_2,\tau_3\sigma_3)$\cite{Tang2016,Wang2017_2}. Among the available Gamma matrices, we may introduce $\tau_3\sigma_3$ to gap out the spectrum of the Hamiltonian in Eq.~(\ref{eq:Hlow}). However, such terms are forbidden by the gliding mirror reflection symmetry $\mathcal{G}_x=i\tau_3\sigma_1$ and, therefore, the Dirac points are protected by the additional non-symmorphic symmetry $\mathcal{G}_x$.
Once the N\ee{l} vector deviates from $[100]$, $\mathcal{G}_x$ is broken and the low-energy effective model develops a gap with a magnitude proportional to the staggered Zeeman energy $\Delta$ following the details in Appendix~\ref{sec:Heff}. To make a clear distinction between gapless and gapped energy spectra, a large Zeeman energy $\Delta$ is desirable. However, we find in Eq.~(\ref{eq:DeltaCond}) that the Zeeman energy needs to be smaller than the SOC strength to ensure the existence of the Dirac points. Therefore, our model requires a strong SOC to ensure the Dirac points while a Zeeman energy comparable to the SOC strength is desirable for a clear distinction between gapped and gapless phase.

We now confirm the above-mentioned analysis by numerically evaluating the tight-binding Hamiltonian in Eq.~(\ref{eq:tb2}).
First of all, we examine the eigenvalue spectrum for different N\ee{l} vector configurations to observe symmetry protected Dirac points and corresponding gapless phase.
Fig.~\ref{fig:spectrum}(a) shows the eigenvalue spectrum with the following set of parameters: $t_{xy}=t=0.5$ eV, $t_{z}=0.5t$, $t'_{xy}=0.1t$, $t'_z=0.1t$, $\lambda=0.5t$, $\lambda_z=0.1t$, and $\Delta=0.3t$. We choose these parameters to ensure two DNLs in the system and a clear gap exists when non-symmorphic symmetry is broken. The eigenvalue spectrum is plotted for the three different N\'{e}el vector configurations, which are the $\hat{\bm{n}}||[100]$ (black solid line), $\hat{\bm{n}}||[110]$ (blue dashed line), and $\hat{\bm{n}}||[001]$ (red dotted line). Among the three different configurations, the $\hat{\bm{n}}||[100]$ (black solid line) configuration preserves $\mathcal{G}_x$ and the Dirac points are protected by the non-symmorphic symmetry at $k_x=\pi$ along $X-M$ and $R-L$ lines. Similarly, the $\hat{\bm{n}}||[010]$ configuration also preserves $\mathcal{G}_y$, and the Dirac points are protected along $k_y=\pi$, but it is not shown here. Otherwise, the Dirac points are gapped as shown in the $\hat{\bm{n}}||[110]$ and $\hat{\bm{n}}||[001]$ configurations in Fig.~\ref{fig:spectrum}(a). 
In addition, we vary the staggered Zeeman energy $\Delta$ to examine the number of allowed DNLs according to the conditions in Eq.~(\ref{eq:DeltaCond}). 
Fig.~\ref{fig:spectrum}(b) shows the eigenvalue spectrum with the same parameters used in Fig.~\ref{fig:spectrum}(a) with $\hat{\bm{n}}||[100]$, but different staggered Zeeman energy $\Delta$. The black solid line shows the system parameters with $\Delta=0.3t$, which satisfies the condition $\Delta<\lambda-\lambda_z$. As previously discussed in Eq.~(\ref{eq:DeltaCond}), two DNLs exist for this parameter and, consequently, Fig.~\ref{fig:spectrum}(c) shows two dispersive DNLs in $\hat{k}_y-\hat{k}_z$ plane at $k_x=\pi$, which both cross $X-M$ and $R-L$ lines. The blue dotted line in Fig.~\ref{fig:spectrum}(b) depicts the spectrum with $\Delta=0.5t$, where the parameters satisfy the condition $\lambda-\lambda_z<\Delta<\lambda+\lambda_z$ and the system has one DNL. Fig.~\ref{fig:spectrum}(d) clearly shows one DNL which crosses $R-L$ line, but not $X-M$ line. The gray dashed line in Fig.~\ref{fig:spectrum}(b) has parameter $\Delta=1.0t$ which satisfies $\Delta>\lambda+\lambda_z$. Therefore, the spectrum is fully gapped and the system has no Dirac points even though $\mathcal{G}_x$ is not broken.

\section{Magnetic anisotropy energy in Antiferromagnetic Dirac semimetal}\label{sec:Free}

With an established model, we now wish to explore the effects of electrostatic gating on the resultant magnetization of the AFS.
In the following sections, we analyze the underlying physics of the voltage driven N\ee{l} vector manipulation.

\subsection{Free energy analysis} \label{sec:FEA}
Voltage induced magnetic order control has been proposed in a 3D topological insulator-ferromagnetic insulator (TI-FI) hybrid system\cite{Kiwook2012}. In TI-FI hybrid system, a finite magnitude of exchange energy is induced by the proximity effect at the surface of the TI. In this case, the electronic structure of the TI surface state may be affected by the induced magnetic order. When the magnetic order in FI is perpendicular to the TI surface, the induced exchange energy causes an energy gap in TI surface states. In contrast, the surface state is still gapless if the magnetic order of FI is parallel to the TI surface\cite{Qi2011}. In this case, the induced exchange coupling merely shifts the Dirac point of the TI surface state in momentum space. 
The perpendicular and parallel configuration of the magnetic order results in a gapped and gapless electronic states, making a difference in the total energy of the system. 

Similarly, the electronic structure of AFS is gapless when the N\ee{l} vector is parallel to the mirror reflection symmetry axis (e.g. $\hat{\bm{n}}||[100]$), whereas the spectrum is gapped when the N\ee{l} vector is tilted away from the symmetry axis (e.g. $\hat{\bm{n}}||[001]$). 
To determine the preferred configuration, we evaluate the free energy of the system. 
In other words, we assume that the magnetic easy-axis is aligned with the N\ee{l} vector orientation that minimizes the electron free energy for a given AFS system\cite{Kiwook2012}. 
The Helmholtz free energy, $F$, for fermions is\cite{Pathria1996}
\begin{equation} \label{eq:F}
\frac{F(\varphi,\theta,\mu)}{V}=-\frac{1}{\beta}\frac{1}{(2\pi)^3}\sum_{i}\int d^3k \ln\left( 1+e^{\beta [\mu-E_{\mathbf{k},i}(\varphi,\theta)]} \right),
\end{equation}   
where $\beta=1/k_B T$, $T$ is temperature, $k_B$ is Boltzmann constant, $E_{\bm{k},i}(\varphi,\theta)$ is the eigenvalue of the $i$-th band at $\bm{k}$ when the N\ee{l} vector is $\hat{\bm{n}}=(\cos\varphi\cos\theta,\sin\varphi\cos\theta,\sin\theta)$ with in-plane ($\varphi$) and out-of-plane ($\theta$) angle, $\mu$ is the chemical potential, and $V$ is the volume of the system (see Appendix~\ref{sec:FE} for the derivation). 

We are particularly interested in the free energy difference between various N\ee{l} vector orientations and the corresponding AFS eigenvalue spectrum. 
 \begin{figure}[t!] 
  \centering
   \includegraphics[width=0.5\textwidth]{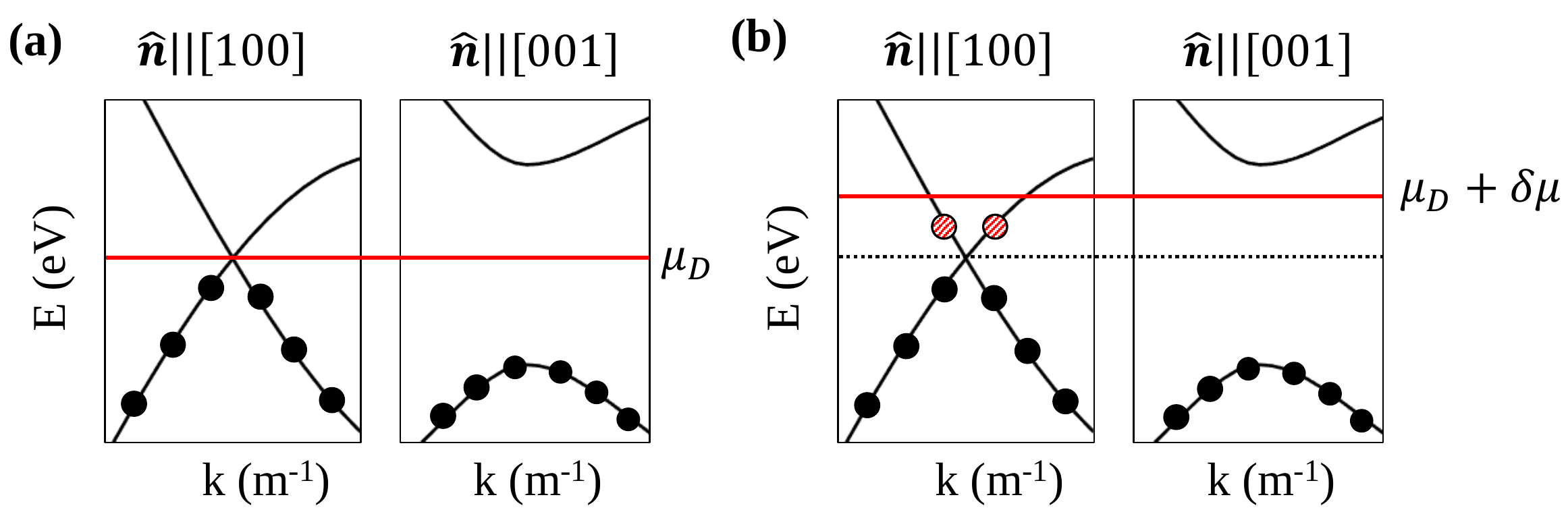}
  \caption{(a) A schematic eigenvalue spectrum near the Dirac point. The left hand side shows the gapless phase whose Dirac point is protected by the $\mathcal{G}_x$. The right hand side shows the gapped spectrum due to the broken gliding symmetry. The filled circle describes the filled electronic states and the chemical potential is located at the Dirac point $\mu_D$. (b) The same plot as (a), but the chemical potential is away from the Dirac point by $\delta\mu$. The gapless phase has additional electron filling indicated by patterned red circles, which provide additional negative energy and lower the total free energy. }\label{fig:bschematic}
\end{figure}   
The Fig.~\ref{fig:bschematic}(a) and (b) show the schematics of the spectrum for two different N\ee{l} vector configurations: $\hat{\bm{n}}||[100]$ (or $\theta=0$ and $\varphi=0$, gapless) and $\hat{\bm{n}}||[001]$ ($\theta=\frac{\pi}{2}$, gapped). In each of these figures, the filled circles indicate the filled electronic states. When the chemical potential is located at the Dirac point $\mu_D$, as in Fig.~\ref{fig:bschematic}(a), the gapped spectrum has lower total energy and, according to Eq.~(\ref{eq:F}), lower free energy. Therefore, $\hat{\bm{n}}||[001]$ configuration is preferred. However, as the chemical potential moves away from the Dirac point by $\delta\mu>0$, more electronic states are filled for gapless spectrum, as shown in Fig.~\ref{fig:bschematic}(b). The additional filling of these electronic states lowers the free energy for gapless spectrum mitigating the initial free energy difference. When the chemical potential is closer to the valence band ($\delta\mu<0$), the overall increase in valence band electron energy compensates the initial free energy difference between the two configurations. Although gapped spectrum is preferred when the chemical potential is located at the Dirac point, the energy difference that favors one phase may become less as the chemical potential moves away from the Dirac point. Similar trend has been found in ferromagnetic material proximity coupled to the graphene\cite{Kiwook2008} and TI surface states\cite{Kiwook2012,Kiwook2017}.

\subsection{Intrinsic anisotropy energy and the chemical potential induced N\ee{l} vector switching} \label{sec:NA}

 \begin{figure}[t!] 
  \centering
   \includegraphics[width=0.45\textwidth]{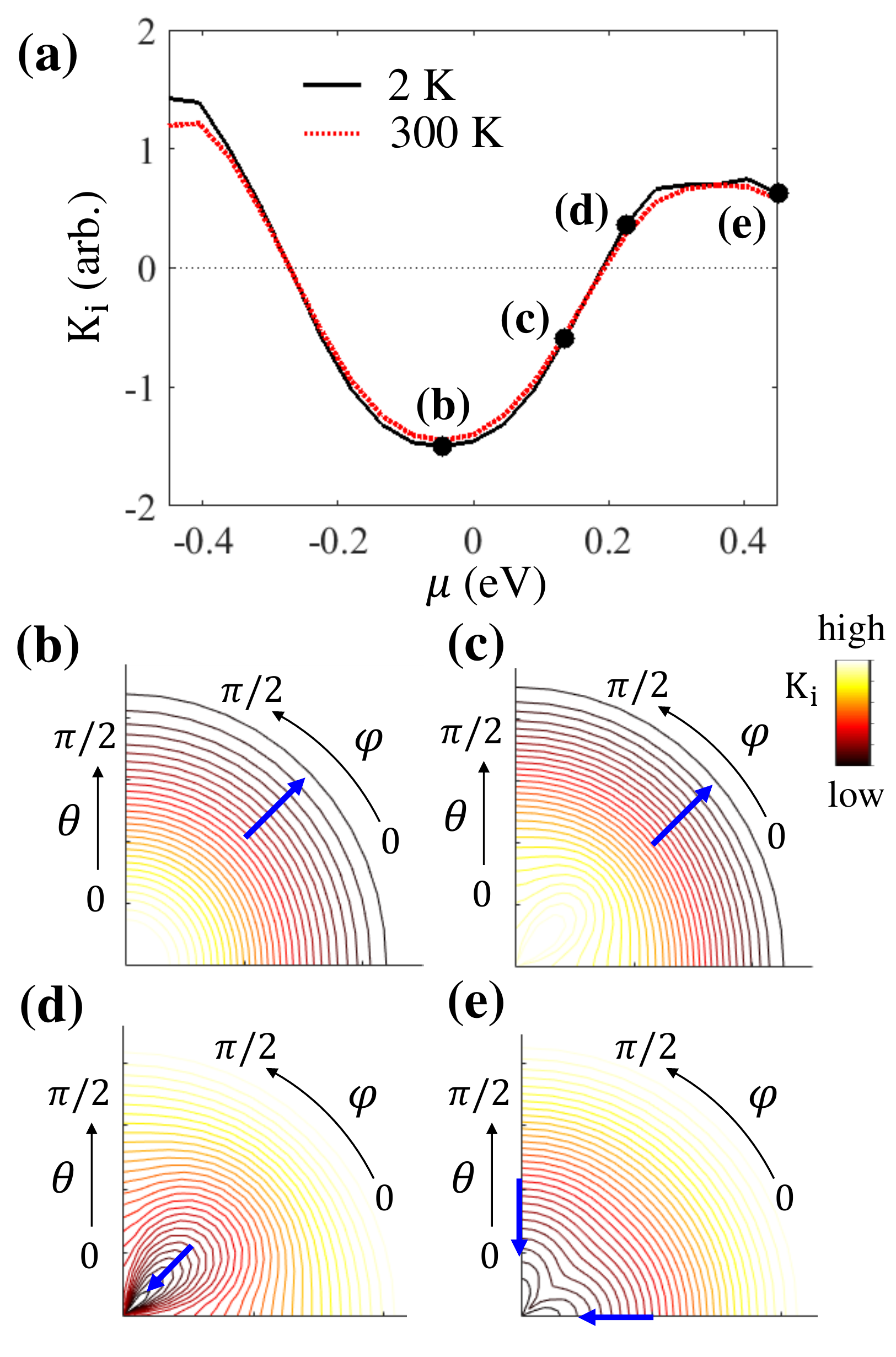}
  \caption{(a) The anisotropy energy $K_i$ as a function of the chemical potential $\mu$. $K_i$ measures the free energy difference between $\hat{\bm{n}}||[001]$ and $\hat{\bm{n}}||[100]$ configuration. When the $K_i$ is negative value, the system prefers $\hat{\bm{n}}||[001]$ configuration (gapped) whereas the preferred N\ee{l} vector configuration is $\hat{\bm{n}}||[100]$ for the system with positive $K_i$. $K_i$ is calculated for $T=2$ K (black solid line) and $T=300$ K (red dotted line). The black dot shows the sampling point of the anisotropy energy landscape shown in (b)-(e). (b) The anisotropy energy landscape for $\varphi\in[0,\pi/2]$ and $\theta\in[0,\pi/2]$ at $\mu\simeq-0.045$ eV. (c) Similar plot as (b) at $\mu\simeq0.135$ eV, (d) $\mu\simeq0.225$ eV, and (e) $\mu\simeq0.45$ eV. The blue arrow indicates the global minimum points. }\label{fig:FE}
\end{figure}   

To examine the scenario described above, we define the intrinsic anisotropy energy\cite{Kiwook2012} of the system as 
\begin{equation} \label{eq:Ki}
K_i(\varphi,\theta,\mu)=\frac{F(\varphi,\theta,\mu)-F(0,0,\mu)}{V},
\end{equation}
and numerically find $\hat{\bm{n}}$ which minimizes $K_i(\varphi,\theta,\mu)$. 
We solve the free energy by using the tight-binding Hamiltonian in Eq.~(\ref{eq:tb2}) with the parameters used in Fig.~\ref{fig:spectrum}(a). We vary both in-plane ($\varphi$) and out-of-plane ($\theta$) angle of the N\ee{l} vector. 
The result shows no substantial change for different temperature, $T$, unless $T$ is comparable to the energy gap of the gapped phase smearing out the energy difference between gapped and gapless phase. From Fig.~\ref{fig:spectrum}(a), we observe that $\hat{\bm{n}}||[001]$ shows the energy gap of $E_g\simeq0.2$ eV, whereas $\hat{\bm{n}}||[110]$ has $E_g\simeq0.12$ eV.
Therefore, the results are insensitive to the range of temperature from $T=2$ K to $T=300$ K. We then obtain the anisotropy energy by evaluating Eqs.~(\ref{eq:F}) and (\ref{eq:Phi}). When the chemical potential is located near the Dirac points ($\mu\simeq0$), the gapped spectrum shows lower free energy. Thus, the system prefers the N\ee{l} vector configuration $\hat{\bm{n}}||[001]$ that induces the largest gap in spectrum. Fig.~\ref{fig:FE}(a) quantitatively describes such trend where we plot the anisotropy energy $K_i(0,\frac{\pi}{2},\mu)$, as a function of chemical potential. For simplicity in notation, we define
\begin{equation} \label{eq:Phi}
K_i(\mu)\equiv K_i(0,\frac{\pi}{2},\mu),
\end{equation}
which measures the free energy difference between maximally gapped ($\hat{\bm{n}}||[001]$ or $\theta=\frac{\pi}{2}$) and gapless ($\hat{\bm{n}}||[100]$ or $\theta=0$ and $\varphi=0$) phase, and we omit $(\varphi,\theta)=(0,\frac{\pi}{2})$ for notational simplicity unless otherwise mentioned. 
As we move the chemical potential toward the conduction band from the Dirac point, gapless spectrum may facilitate more filled states whereas gapped spectrum has no additional occupations in zero temperature limit. The additional filled states in the gapless spectrum further lower the free energy, reducing the initial free energy difference $K_i(\mu)$. After the chemical potential crosses the energy gap, the sign of the anisotropy energy is reversed and $\hat{\bm{n}}||[100]$ is preferred. To analyze the results in more detail, we sample four different points from Fig.~\ref{fig:FE}(a) and plot the free energy landscape for $\varphi\in[0,\pi/2]$ and $\theta\in[0,\pi/2]$ in Fig.~\ref{fig:FE}(b)-(e). In the contour plot, larger (smaller) value of $K_i$ is indicated as light yellow (dark red) color. The $\hat{x}-\hat{y}$ axis represents the out-of-plane angle $\theta$, whereas the radial angle represents the in-plane angle $\varphi$. Fig.~\ref{fig:FE}(b) shows that the free energy minima occur at $\theta=\frac{\pi}{2}$. The anisotropy energy magnitude is decreased but still shows the same minima in Fig.~\ref{fig:FE}(c). Further increase in chemical potential eventually changes the global minima from $\theta=\pi/2$ to $\theta=0$ as it is shown in Fig.~\ref{fig:FE}(d). Note that the global minima is still in gapped phase, but with a smaller gap at $\varphi=\pi/4$, or $\hat{\bm{n}}||[110]$. Eventually, Fig.~\ref{fig:FE}(e) shows that the global minima is located along the $\hat{x}$ ($\varphi=0$) or $\hat{y}$ ($\varphi=\pi/2$) axis as we further tune the chemical potential away from the Dirac point. In summary, Fig.~\ref{fig:FE}(b)-(e) shows a gradual change of the preferred N\ee{l} vector configuration from $\hat{\bm{n}}$ that gives us a maximally gapped spectrum to gapless spectrum via manipulation of the chemical potential. 

 \begin{figure}[t!] 
  \centering
   \includegraphics[width=0.5\textwidth]{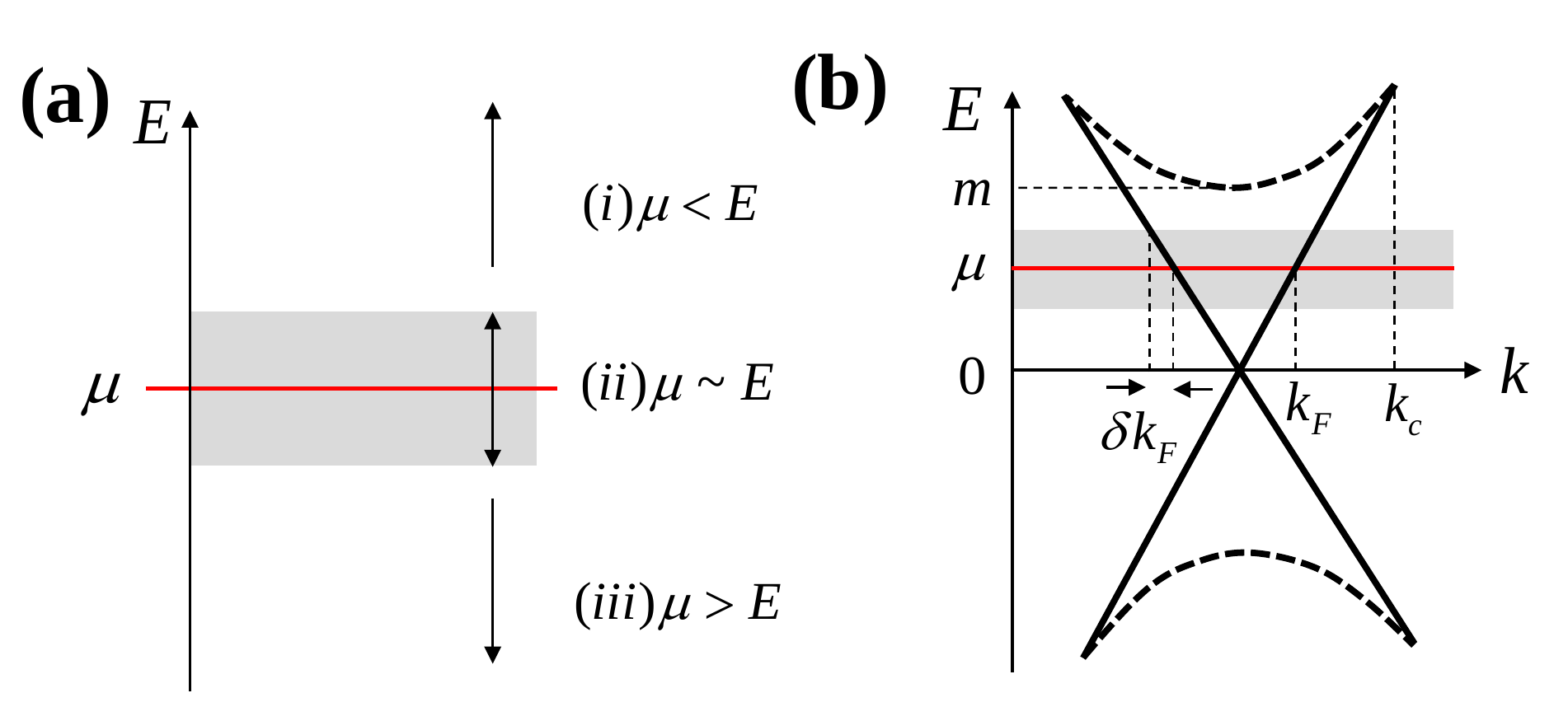}
  \caption{(a) The three regions of interest for linearizing the free energy in Eq.~(\ref{eq:f}). Assuming the zero-temperature limit, we divide the region when the energy of interest is in (i) $\beta(\mu-E)\ll -1$, (ii) $\beta|\mu-E|\ll 1$, and (iii) $\beta(\mu-E) \gg1$. (b) A schematics of the low-energy Dirac fermion with an isotropic Fermi velocity. The Dirac point is located at $E=0$. The solid line shows the gapless, linear dispersion of the massless Dirac fermion. The dashed line depicts the gapped spectrum of the massive Dirac fermion with a mass term $m$. The shaded region indicates the region where $\beta|\mu-E|\ll 1$ condition is satisfied and the corresponding momentum range is $k_F-\delta k_F<k<k_F+\delta k_F$.  }\label{fig:Fmu}
\end{figure}   

\subsection{Intrinsic anisotropy energy dependency on the model Hamiltonian parameters} \label{sec:AA}
In Section~\ref{sec:NA}, we examine the intrinsic anisotropy energy as a function of the chemical potential, and observe the voltage induced N\ee{l} vector switching. 
However, it is unclear how the specific magnitude of $K_i$ and the decreasing rate of its magnitude are determined.   
In this regard, we linearize Eq.~(\ref{eq:F}) and perform a closer analysis on the free energy to analyze parametric dependencies of $K_i$.  
The free energy in Eq.~(\ref{eq:F}) is rewritten using $F/V\propto \sum_i \int d^3k f_{\vec{k},i}$ where 
\begin{equation} \label{eq:f}
f_{\vec{k},i}=-\frac{1}{\beta}\ln \left( 1+e^{\beta(\mu-E_{\vec{k},i})} \right).
\end{equation}
Eq.~(\ref{eq:f}) shows the individual eigenstate contribution to the free energy, which may be linearized in zero-temperature limit $T\rightarrow 0$, or $\beta\rightarrow\infty$. Specifically, we consider the three different energy ranges shown in Fig.~\ref{fig:Fmu}(a). The linearized $f_{i,\vec{k}}$ for each of the three energy ranges are derived in Appendix~\ref{sec:FElin} in the zero-temperature limit, and we summarize the results as follows:
\begin{equation} \label{eq:flin}
f_{\vec{k},i}\simeq
\begin{cases} 
0 & \text{if } \mu<E_{\vec{k},i}, \\
-(\mu-E_{\vec{k},i})/2 & \text{if } \mu\sim E_{\vec{k},i}, \\
-(\mu-E_{\vec{k},i}) & \text{if } \mu > E_{\vec{k},i}. \\
\end{cases}
\end{equation}
To proceed, we consider eigenstates of the 3D Dirac fermions by using massive, isotropic Dirac Hamiltonian. The eigenstates of such Hamiltonian are $E_{k,1}=-\sqrt{(v_F k)^2+m^2}$ for the valence band and $E_{k,2}=\sqrt{(v_F k)^2+m^2}$ for the conduction band, where we set $\hbar=1$, $m$ is the mass term, and $v_F$ is the isotropic Fermi velocity. To simplify our analysis, we ignore the two-fold degeneracy of the Dirac Hamiltonian, but the below analysis is physically valid, as degeneracy plays no role but simply adds a factor of two to the total free energy. 
We restrict the location of the chemical potential within the gap of the gapped phase, as it gives a simple, yet useful solution to examine the free energy change as a function of the chemical potential deviation from the Dirac point.

We first compute the free energy of the gapless phase ($m=0$) for $\mu\geq0$ by solving Eq.~(\ref{eq:F}) in conjunction with Eq.~(\ref{eq:flin}). Fig.~\ref{fig:Fmu}(b) depicts a particular momentum cut of the massless (solid line) and massive (dashed line) Dirac fermion dispersions. The shaded region in Fig.~\ref{fig:Fmu}(b) shows a range of energy that is close to the chemical potential and satisfies $f_{i,\vec{k}}\simeq-(\mu-E_{i,\vec{k}})/2$ in Eq.~(\ref{eq:flin}).
Then, we obtain the free energy by integrating over the relevant energy range as
\begin{equation} \label{eq:Fgapless}
\begin{split}
\frac{F(m=0)}{V}\propto& -\int_0^{k_c} k^2 dk (\mu+v_F k) \\
&-\int_0^{k_F-\delta k_F} k^2 dk (\mu-v_F k) \\
&-\int_{k_F-\delta k_F}^{k_F+\delta k_F} k^2 dk \frac{\mu-v_F k}{2},
\end{split}
\end{equation}
where we assume that the low-energy Dirac Hamiltonian is a valid description for $k\leq k_c$ with an arbitrary cut-off wavevector $k_c$, and $\delta k$ is an infinitesimally small deviation from the Fermi wavevector $k_F=\mu/v_F$. In Eq.~(\ref{eq:Fgapless}), the first term accounts for the filled states in the valence band, the second term considers the filled states in the conduction band, and the last term includes the states near the Fermi level, where $\delta k_F$ is determined by the temperature. In the zero-temperature limit, $\delta k_F\rightarrow 0$ and Eq.~(\ref{eq:Fgapless}) becomes
\begin{equation}
\frac{F(m=0)}{V}\propto -\left( \frac{\mu}{3} + \frac{v_F k_c}{4} \right)k_c^3-\frac{1}{12}\frac{\mu^4}{v_F^3}.
\end{equation} 
We now consider the free energy of the gapped phase with a mass term $m$ for $0\leq \mu < m$. The free energy is
\begin{equation} \label{eq:Fgapped}
\frac{F(m)}{V}\propto  
-\int_0^{k_c} k^2 dk (\mu+\sqrt{(v_F k)^2+m^2}),
\end{equation} 
as we only need to consider the filled states in the valence band.
Here, we assume that our cut-off wavevector is sufficiently large satisfying $v_F k_c \gg m$. In this case, $x=v_Fk_c/m\gg1$ and we use the limit of the integration $\int_0^x dx' {x'}^2\sqrt{{x'}^2+1}\simeq x^4/4+x^2/4$ for $x\rightarrow \infty$ to evaluate Eq.~(\ref{eq:Fgapped}). As a result, we obtain
\begin{equation}
\frac{F(m)}{V}\propto 
-\left( \frac{\mu}{3} + \frac{v_Fk_c}{4} \right)k_c^3
-\frac{m^2}{4v_F}k_c^2.
\end{equation}
Finally, we obtain the zero-temperature limit intrinsic anisotropy energy by using the definition in Eq.~(\ref{eq:Phi}) as follows:
\begin{equation} \label{eq:Kimu}
\begin{split}
K_i(\mu)=&\frac{F(m)-F(m=0)}{V} \\
\propto&-\frac{1}{4v_F^3}\left( m^2 (v_F k_c)^2-\frac{\mu^4}{3} \right).
\end{split}
\end{equation}
Following the same procedures outlined in Eqs.~(\ref{eq:Fgapless})-(\ref{eq:Kimu}), we obtain identical results for $-m<\mu\leq0$. Therefore, Eq.~(\ref{eq:Kimu}) is valid for $-m<\mu<m$ in the zero-temperature limit. Eq.~(\ref{eq:Kimu}) shows that the magnitude of the intrinsic anisotropy is proportional to $m^2$ and inversely proportional to $v_F$ at $\mu=0$. In other words, a large $m$ and small $v_F$ is desirable to maximize the intrinsic anisotropy energy. For an increasing $\mu$, small $m v_F$ is favorable to have a rapid decrease in the magnitude of the $K_i$. This may result in a sign change of the $K_i$ and induce a switching of the preferred N\ee{l} vector orientation. For this reason, a large gap, $m$, and a small Fermi velocity, $v_F$, is desirable to realize the voltage driven N\ee{l} vector switching.  

 \begin{figure}[t!] 
  \centering
   \includegraphics[width=0.5\textwidth]{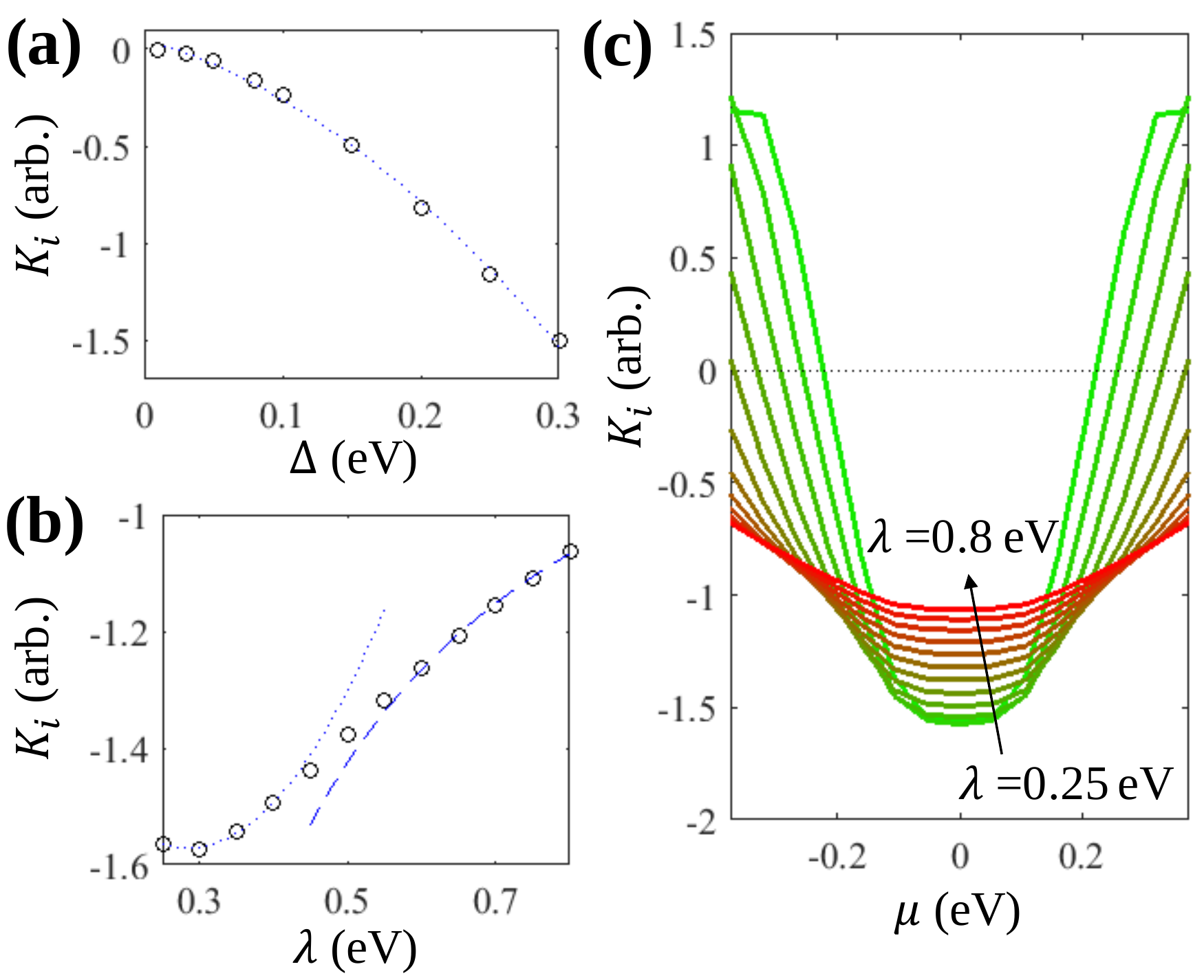}
  \caption{(a) The intrinsic anisotropy energy, $K_i$, at $\mu=0$ as a function of the Zeeman energy $\Delta$. The dotted line shows a quadratic fitting showing a good agreement with Eq.~(\ref{eq:KiD}). (b) $K_i$ at $\mu=0$ as a function of the SOC strength $\lambda$. The dotted line shows a quadratic fitting and dashed line shows a fitting curve with $1/\lambda$. The two distinct limit shows the dependence of $K_i$ on $\lambda$ as it is discussed in Eq.~(\ref{eq:KiL}). (c) $K_i$ as a function of the chemical potential, $\mu$, for various $\lambda$. The green line represents a small $v_F$ limit, or small $\lambda$ limit whose size is comparable to $\Delta$. The red line represents a large $v_F$ limit, or large $\lambda$ limit.    }\label{fig:analytic}
\end{figure}   
We now wish to analyze how the low-energy Dirac fermion parameters, $m$ and $v_F$, are expressed in terms of the parameters in our model Hamiltonian. To this end, we first consider the low-energy Hamiltonian near the Dirac point. In Section~\ref{sec:TBK}, we find that the energy gap of the low-energy Hamiltonian is proportional to the staggered Zeeman energy $\Delta$. Therefore, the mass term of the Dirac fermion is proportional to the Zeeman energy, or $m\propto \Delta$. According to Eq.~(\ref{eq:Kimu}), the intrinsic anisotropy energy satisfies $K_i(0)\propto -m^2$ and, as a result, 
\begin{equation} \label{eq:KiD}
K_i(0)\propto -\Delta ^2,
\end{equation}
for a fixed Fermi velocity $v_F$. To numerically show such dependency, we obtain $K_i(0)$ by evaluating Eq.~(\ref{eq:Phi}) using the same parameters in Fig.~\ref{fig:spectrum}(a), but we set the second-nearest neighbor hopping to zero, or $t_{xy}'=t_{z}'=0$, in order to locate the Dirac points at $E=0$. To evaluate the anisotropy energy in zero-temperature limit, we set the temperature $T=2$ K. Fig.~\ref{fig:analytic}(a) shows $K_i(0)$ as a function of $\Delta$ from $300$ meV to $5$ meV, and the dashed line is a quadratic fitting of the numerical results showing that the intrinsic anisotropy energy satisfies Eq.~(\ref{eq:KiD}). As a result, we may increase the intrinsic anisotropy energy, $K_i$, by introducing a large Zeeman energy. This is due to the fact that $K_i$ is proportional to the difference between the total energy of the gapped and gapless spectrum, which is determined by the size of the gap.

For a fixed Zeeman energy, Eq.~(\ref{eq:Kimu}) shows that $K_i(0)$ is inversely proportional to the Fermi velocity $v_F$. Thus we wish to identify a relevant parameter within our model Hamiltonian that corresponds to $v_F$. 
To this end, we examine the low-energy spectrum of Eq.~(\ref{eq:Hlow}),
\begin{equation} \label{eq:Eq2}
E(\vec{k}_1+\vec{q})=\pm\sqrt{v_x^2 q_x^2 + (v_y q_y + v_z q_z)^2},
\end{equation}
where $v_x^2=\sqrt{v_{x1}^2+v_{x2}^2+v_{x3}^2}$, and $v_{x1},v_{x2},v_{x3},v_y,v_z$ are given in Eq.~(\ref{eq:vxyz}). Assuming that spin-orbit strength of the layered structure is dominantly determined by the in-plane spin-orbit coupling strength $\lambda$, we further simplify Eq.~(\ref{eq:Eq2}) by ignoring $\lambda_z$, or setting $\lambda_z=0$. Eq.~(\ref{eq:Eq2}) is then simplified as $E(\vec{k}_1+\vec{q})=\pm\sqrt{v_x^2 q_x + v_y^2 q_y^2}$, where $v_x^2=v_{x12}^2+\lambda^2$ with $v_{x12}^2=v_{x1}^2+v_{x2}^2=\frac{1+\cos k_{y0}}{8}(t_{xy}^2+t_z^2+2t_{xy}t_z\cos k_{z0})$, $v_y=-\lambda\cos k_{y0}$, and $\sin k_{y0}=\Delta/\lambda$. In large SOC limit, or $\lambda \gg \{\Delta,t_{xy}, t_z\}$, $v_x^2=v_y^2\simeq \lambda^2$ and the low-energy spectrum follows $E\simeq \pm \lambda \sqrt{q_x^2+q_y^2}$ with the Fermi velocity $v_F\simeq\lambda$. Note that the Dirac points exist when $\lambda\geq \Delta$ according to Eq.~(\ref{eq:DeltaCond}). Therefore, the smallest $\lambda$ we may consider is $\lambda\sim\Delta$. In this limit, $\cos k_{y0}=\sqrt{1-(\Delta/\lambda)^2}\simeq0$ and $v_y=-\lambda\cos k_{y0}\simeq 0$. Then the low-energy spectrum becomes $E\simeq v_x q_x$ with the Fermi velocity $v_F=v_x=\sqrt{v_{x12}^2+\lambda^2}$.
By considering the above mentioned relationship between $v_F$ and $\lambda$, we show the $K_i(0)$ dependency on $\lambda$ for two different limits as follows:
\begin{equation} \label{eq:KiL}
\begin{split}
&K_i(0)\propto -\frac{1}{v_F}  \\
&\simeq
\begin{cases}
-\frac{1}{v_x}\simeq -\frac{1}{v_{x12}}+\frac{1}{2v_{x12}^3}\lambda^2 & \text{if } \lambda\sim\Delta\text{ and }\lambda\ll v_{x12}, \\
-\frac{1}{\lambda} & \text{if }\lambda\gg \{v_{x12},\Delta \}, \\
\end{cases}
\end{split}
\end{equation}
for a fixed Zeeman energy $\Delta$, and $\lambda_z=0$. 
We numerically verify Eq.~(\ref{eq:KiL}) by evaluating the intrinsic anisotropy energy of the model Hamiltonian using Eq.~(\ref{eq:Phi}). We use the same parameters as used in Fig.~\ref{fig:spectrum}(a) but here set $\lambda_z=0$ following the assumption made in Eq.~(\ref{eq:KiL}), and set $t_{xy}'=t_{z}'=0$ to locate all the Dirac points at $E=0$. 
Fig.~\ref{fig:analytic}(b) shows $K_i(0)$ as a function of $\lambda$ from $0.25$ eV to $0.8$ eV. Note that $v_{x12}\simeq 0.375$ eV is the maximum value of $v_{x12}$, thus we expect to see $K_i(0)\propto \lambda^2$ near $0.25$ eV (dotted line) and $K_i(0)\propto-1/\lambda$ near $0.8$ eV (dashed line). Consequently, the overall trend shows that $K_i(0)$ is a decreasing function for increasing $\lambda$, and this is due to the fact that $\lambda$ sets the Fermi velocity in our model. The linear dispersion of the low-energy Dirac fermions possesses a steeper slope for larger $v_F$ and, therefore, there are few available states for a given range of energy. The reduced number of filled states results in a smaller free energy difference between gapped and gapless phase, or smaller $K_i(0)$. 

The above scenario is directly related to the behavior of $K_i$ as a function of the chemical potential change. As we increase the chemical potential from the Dirac point, more states are filled for the gapless phase lowering the total free energy while the number of filled states remain the same for gapped phase. If we increase $v_F$, the slope of the linear dispersion becomes steeper and the density of states become smaller. Then fewer states are included upon an increase of the chemical potential and, as a result, a change in the total free energy for gapless phase becomes smaller. This trend is not desirable as we wish to observe a rapid decrease in the total free energy of the gapless phase as we increase $\mu$ in order to realize a switching of the preferred spectrum from gapped to gapless. In this regard, small Fermi velocity is preferred and, equivalently, small SOC strength, $\lambda$, is desirable. Note that $\lambda$ still needs to be larger than $\Delta$ in order to ensure the existance of the symmetry protected Dirac points according to Eq.~(\ref{eq:DeltaCond}), therefore, the smallest possible SOC strength is $\lambda\sim\Delta$. 
Fig.~\ref{fig:analytic}(c) shows the calculated $K_i(\mu)$ as a function of $\mu$ using the same parameters in Fig.~\ref{fig:analytic}(b). $K_i$ shows a clear sign change at $\mu\simeq 0.17$ eV for $\lambda=0.25$ eV, which corresponds to a small $v_F$ limit. By contrast, we no longer observe a sign change in $K_i(\mu)$ for $\lambda=0.8$ eV within a given range of $\mu$, which corresponds to the large $v_F$ limit.

\begin{figure*}
  \centering
   \includegraphics[width=1.0\textwidth]{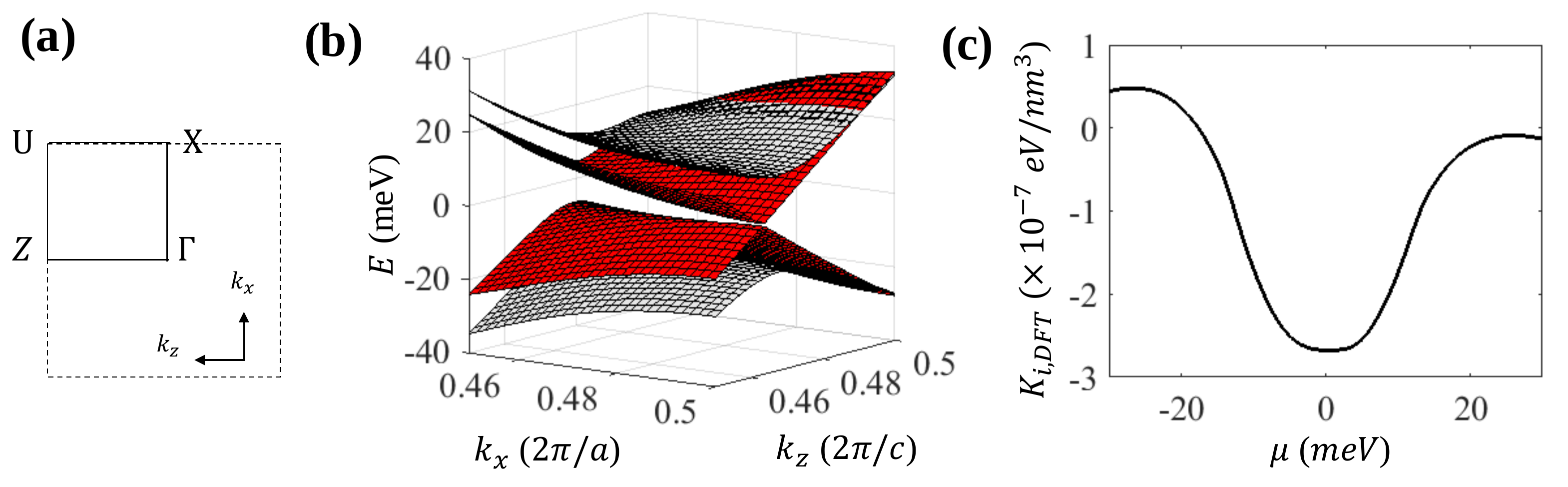}
  \caption{(a) The Brillouin zone of orthorhombic CuMnAs at $k_y=0$. (b) The eigenvalue spectrum obtained from DFT calculation near the protected Dirac point. The red plot represents the spectrum obtained from $\hat{\bf{n}}||[001]$ configuration, which clearly shows the Dirac point at $k_x=\pi/a$. The grey plot shows the gapped spectrum in $\hat{\bf{n}}||[100]$ configuration. (c) A plot of the intrinsic anisotropy energy, $K_{i}$, calculated by using DFT results. The $K_i$ is calculated using Eq.~(\ref{eq:Ki}) using $dk^3=(2\pi)^2/(abc)$ where the lattice constants in $x$, $y$, $z$ are $a=6.439$ \AA, $b=3.800$ \AA, $c=7.292$ \AA, respectively. 
   }\label{fig:DFT}
\end{figure*}   
\subsection{Intrinsic anisotropy energy in orthorhombic CuMnAs} \label{sec:DFT}

In our model, we considered the low-energy Hamiltonian which only captures the eigenvalues near the Fermi level. In addition, our model only contains symmetry protected Dirac fermions whose gapped and gapless phase are governed by the N\ee{l} vector orientation. However, in a realistic material, multiple bands are available, and the system may have accidental crossings of the bands, which are not protected, thus possess an arbitrary gap size. To address this, we examine the intrinsic anisotropy energy of a realistic material using density functional theory (DFT). 

We consider orthorhombic CuMnAs, which is known to be an AFS posessing Dirac fermions located near the Fermi level\cite{Tang2016, Jungwirth2017}. We relax all atomic positions and the unit cell using a $\Gamma$-centered $8\times16\times8$ $k$-point grid. The resulting relaxed lattice parameters are $a=6.439$ \AA, $b=3.800$ \AA, and $c=7.292$ \AA. In this system, the Dirac points are protected by the non-symmorphic two-fold screw symmetry along $z$-axis, $\mathcal{S}_{2z}$, at $k_x=\pi/a$ plane. Before the calculation takes account for SOC, the spectrum possesses Dirac nodal lines on the $k_y=0$ plane\cite{Tang2016} whose high-symmetry points are indicated in Fig.~\ref{fig:DFT}(a). When SOC is included, the spectrum is gapped out and massive Dirac fermions are found along the $\Gamma-X$, $X-U$, and $Z-X$ lines\cite{Tang2016, Jungwirth2017}. When the N\ee{l} vector is aligned with $[001]$, the $\mathcal{S}_{2z}$ symmetry is respected and the protected Dirac point is found in the $X-U$ line.

We first focus on the Dirac point located at $X-U$ line by using $30\times30$ uniform grid for limited range of $k$-space of $0.45(2\pi/a)\leq k_x \leq 0.5(2\pi/a)$ and $0.45(2\pi/c)\leq k_z \leq 0.5(2\pi/c)$ at $k_y=0$. We self-consistently obtain charge density of orthorhombic CuMnAs using $22\times44\times22$ uniformly sampled $k$ grid and perform DFT calculation on $30\times 30$ uniform grid to obtain detailed eigenvalue spectrum near the Dirac cone. 
DFT calculations with the projector-augmented wave (PAW) method for orthorhombic CuMnAs are implemented by Vienna \emph{ab initio} simulation package (VASP)\cite{Kresse1996}. A plane-wave kinetic cutoff is selected as $600$ eV to obtain a convergence of the results. The generalized-gradient approximation by Perdew, Burke and Ernzerhof (PBE)\cite{Perdew1996} is used to describe exchange and correlation. 
The obtained eigenvalues capture the realistic description on the Dirac fermions near the Fermi energy.
The red (dark) surface plot in Fig.~\ref{fig:DFT}(b) clearly shows the protected Dirac point at $k_x=\pi/a$ plane for $\bm{n}||[001]$. In the presence of SOC, $\mathcal{S}_{2z}$ is broken for other N\ee{l} vector configurations. Among other possible N\ee{l} vector configurations, previous calculation\cite{Jungwirth2017} shows that $\bm{n}||[100]$ is the ground state of the system. For this particular N\ee{l} vector configuration, the spectrum is fully gapped having an energy gap of $E_g\simeq 26.5$ meV along $X-U$ line as shown in the gray surface plot in Fig.~\ref{fig:DFT}(b). We then compute the intrinsic anisotropy energy, $K_{i,DFT}$, by using the eigenvalues obtained from DFT results for a given $30\times30$ $k_x-k_z$ grid. Specifically, we calculate 
\begin{equation} \label{eq:KiDFT}
K_{i,DFT}(\mu)=\frac{F_{\bm{n}||[100]}(\mu)-F_{\bm{n}||[001]}(\mu)}{V},
\end{equation}
where $F_{\bm{n}||[001]}(\mu)$ and $F_{\bm{n}||[100]}(\mu)$ are the free energy obtained by Eq.~(\ref{eq:F}) for $\bm{n}||[001]$ and $\bm{n}||[100]$ configuration, respectively. 
We set $T=2$ K when we compute the $K_{i,DFT}$ and we assume an identical initial Fermi level for both N\ee{l} vector configurations. Fig.~\ref{fig:DFT}(c) shows the calculated intrinsic anisotropy energy, and the gapped phase is initially preferred ($\hat{\bm{n}}||[100]$) at $\mu=0$ eV. $K_{i,DFT}$ shows a clear decrease in its magnitude as the chemical potential deviates from the Dirac point. 
In fact, the energy difference between gapped and gapless phase is maintained in wider range of k-space away from our $30\times30$ $k_x-k_z$ grid near the Dirac point, thus the magnitude of $K_{i,DFT}$ is expected to be larger than our results. In other words, we effectively limit our cut-off momentum, $k_c$, in Eq.~(\ref{eq:Kimu}) and, consequently, underestimate $K_{i,DFT}$. In addition, we may find larger $K_{i,DFT}$ by considering a material having a larger energy gap for the gapped phase, as we discussed in Section~\ref{sec:AA}. Nevertheless, we observe a clear decrease in the anisotropy energy for an increasing chemical potential deviation from the Dirac point, and the overall behavior of the $K_{i,DFT}$ is consistent with that of our tight-binding model in Section~\ref{sec:NA}.

To determine the preferred N\'{e}el vector configuration of CuMnAs, we may now perform DFT calculation  for the entire Brillouin-zone and obtain converged total energy. The result shows the total energy difference of $E_{tot,\hat{\bf{n}}||[100]}-E_{tot,\hat{\bf{n}}||[001]}\simeq-0.38$ meV per unit cell, thereby we obtain the gapped spectrum as a ground state spectrum of the material. The result agrees with the previous DFT results\cite{Jungwirth2017} as well as the recent transport measurements where the massive Dirac fermions have been identified in orthorhombic CuMnAs\cite{Ni2017,Lei2017}. However, it is uncertain at this point that our argument in Section~\ref{sec:FEA} may play a major role in determining the ground state configuration of the N\'{e}el vector, as the realistic band structure is complicated. For example, the system may possess accidental linear crossings of the bands which are not protected by any underlying symmetries. In such cases, those particular Dirac fermions acquire a gap for an arbitrary N\ee{l} vector orientation but its size may not be correlated with the N\ee{l} vector. It has been shown that multiple Dirac cones exist along the $\Gamma-X$, $X-U$, and $Z-X$ lines in orthorhombic CuMnAs\cite{Tang2016, Jungwirth2017}. Among them, no non-symmorphic symmetry protects the Dirac points in $\Gamma-X$ and $Z-X$ lines. Therefore, in the presence of SOC, the Dirac cone acquires a mass term both for the $\hat{\bm{n}}||[100]$ and $\hat{\bm{n}}||[001]$ configurations\cite{Tang2016}. However, the Dirac cones along the $\Gamma-X$ and $Z-X$ lines show\cite{Jungwirth2017} larger energy gap for $\hat{\bm{n}}||[001]$ than $\hat{\bm{n}}||[100]$, whereas the Dirac cone in $X-U$ line possesses larger size energy gap for $\hat{\bm{n}}||[100]$ than $\hat{\bm{n}}||[001]$. Due to this opposite trend in energy gap for N\'{e}el vector configurations, we may expect different contribution of each Dirac cones on anisotropy energy. As a result, $K_{i,DFT}$ may show different qualitative behavior from Fig.~\ref{fig:DFT}(c) once the whole BZ is properly considered. Therefore, more comprehensive calculation needs to be done with carefully choosen $k$ grid which samples enough points to identify each of the Dirac cones in order to estimate a quantitative behavior of $K_{i,DFT}(\mu)$ for the realistic material.

\section{Two terminal conductance measurement} \label{sec:current}
\begin{figure}[t!] 
  \centering
   \includegraphics[width=0.5\textwidth]{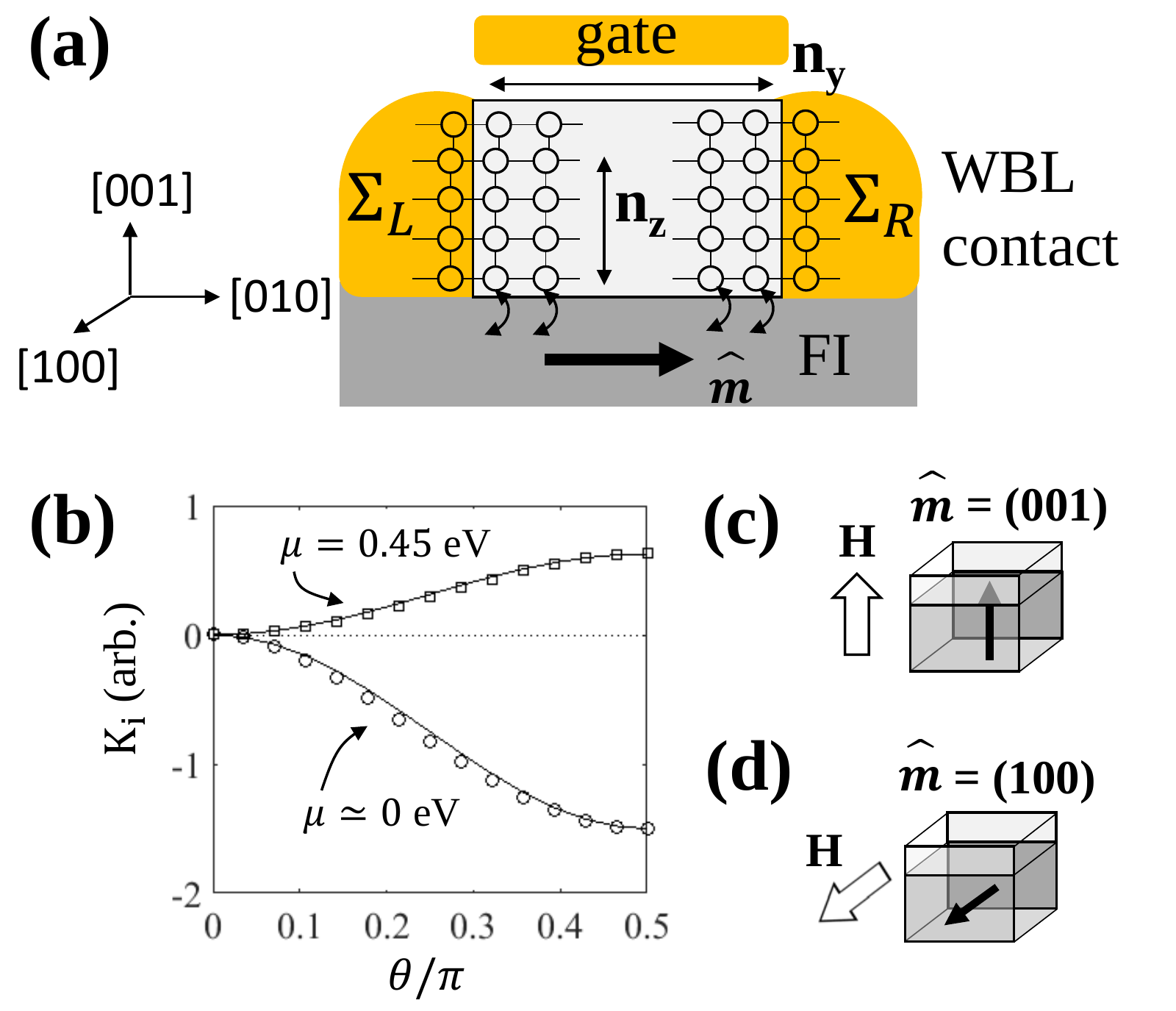}
  \caption{(a) A schematic of the thin slab resistance measurement setup. The AFS material (middle, light gray) is indirectly coupled to the gate, which controls the chemical potential. The contact is directly coupled to the left and right hand side of the AFS and we use wide band limit (WBL) approximation for the metallic contact self-energy. The ferromagnetic insulator (FI) is coupled to the bottom of the AFS and we assume an interface exchange coupling. The net magnetization of the FI is aligned in in-plane direction. (b) The inset shows the anisotropy energy as a function of $\theta$ for fixed $\varphi=0$ when the chemical potential is located at (a) and (e) in Fig.~\ref{fig:FE}(a). The dashed line shows the fit using $\sin^2\theta$. (c-d) Schematics of AFS (light gray) proximity coupled with FI (dark gray). The net magnetization of FI may be changed by applying external field $\mathbf{H}$. 
  }\label{fig:I}
\end{figure}

While it is evident that manipulating the chemical potential changes the orientation of the N\ee{l} vector, thereby driving a topological MIT, the experimental signature of the transition is not yet clear. In light of this, we now explain the signature observed in a two terminal quantum transport measurement within an AFS. 
Fig.~\ref{fig:I}(a) shows the schematics of the AFS which is indirectly coupled to a top-gate to control the chemical potential, and directly connected to the metallic contacts to measure a current across the AFS. However, the conductance alone may not be sufficient to identify the phase of the material. Assuming that the bulk chemical potential is initially at the Dirac point, we may go across the phase transition from insulating to metallic phase as we manipulate the gate voltage. Considering that the phase change occurs near the band edge of the insulating phase, it is unclear whether the increase in conductance is due to the phase transition to the metallic phase or an inclusion of the conduction band edge of the insulating phase. For this reason, we introduce an additional anisotropy energy in the AFS system via exchange coupling with a ferromagnetic insulator (FI). Such induced anisotropy, often referred to as the exchange spring effect, has been investigated for the antiferromagnetic system coupled to the ferromagnetic material\cite{Takahashi2002, Tang2016}. The exchange spring effect allows the external magnetic field to reorient the axis of the induced anisotropy energy, which serves as an additional knob to modify the critical chemical potential at which the phase transition occurs. 
Note that the coupling of FI may introduce a $\mathcal{PT}$ symmetry breaking term at the interface of the AFS-FI system. Although the (semi)metal-insulator transition may still occur in the presence of $\mathcal{PT}$ symmetry breaking terms, the detailed changes in spectrum may be worthwhile to examine as we may have Weyl fermions instead of Dirac fermions in AFS system\cite{Wang2017_1}. However, as our goal is to examine a role of additional anisotropy energy introduced by FI, we ignore any $\mathcal{PT}$ symmetry breaking term in our analysis.  

Figure~\ref{fig:I}(b) shows the anisotropy energy near $\mu=0.045$ eV and at $\mu=0.45$ eV with the same parameter choices as in Fig.~\ref{fig:FE}(a). The uniaxial anisotropy energy is often fitted by $\sin^2\theta$, which is the lowest order approximation in the angle $\theta$ between the magnetic order and magnetic easy axis\cite{Graham2009}. In Fig.~\ref{fig:I}(b), the symbols show the numerical calculation and solid lines show an analytical fit using $\sin^2\theta$, which shows good agreement with numerical results. Using the analytical fit, the intrinsic anisotropy energy is written as
\begin{equation} \label{eq:Keff}
K_i(\varphi,\theta,\mu)=-K_{eff}(\mu)\sin^2\theta,
\end{equation}
where $K_{eff}(\mu)=-K_i(0,\frac{\pi}{2},\mu)$ is plotted in Fig.~\ref{fig:FE}(a). As we observe in Fig.~\ref{fig:FE}(b)-(e), the anisotropy energy is uniformly distributed over the in-plane angle direction. Although we observe $\varphi$ dependency of the anisotropy energy near the critical $\mu$, the difference is much smaller than the difference between in-plane and out-of-plane configurations. Therefore, we ignore the $\varphi$ dependency of the $K_i$ in Eq.~(\ref{eq:Keff}) for simplicity. In the absence of any other additional anisotropy energy, the easy axis is in $\hat{z}$ direction when the chemical potential is initially set to $\mu=0$. However, the $K_{eff}(\mu)$ decreases in magnitude, as shown in Fig.~\ref{fig:FE}(a), until $K_{eff}(\mu_c)=0$ at the critical chemical potential $\mu_c$. When the chemical potential is further increased, the intrinsic anisotropy energy flips its sign and easy-axis becomes the in-plane direction (e.g. $[100]$ direction). 

Although the critical chemical potential $\mu_c$ is determined by the intrinsic band structure of the material, we may alter $\mu_c$ by introducing an additional anisotropy energy. 
In this regard, we now consider the interface exchange coupling energy from the presence of a FI. We assume that the AFS system is exchange coupled to a FI whose net magnetization direction is defined as $\hat{\bm{m}}$. To depict the impact of the exchange coupling on $\mu_c$, we examine two orientations of $\hat{\bm{m}}$: out-of-plane direction $\hat{\bm{m}}=(001)$ and in-plane direction $\hat{\bm{m}}=(100)$ as shown in Fig.~\ref{fig:I}~(c) and (d), respectively.
Each configuration may be realized by applying external magnetic field, denoted as $\bm{H}$ in Fig.~\ref{fig:I} (c) and (d), along the in-plane and out-of-plane directions. The exchange coupling introduces an additional anisotropy energy\cite{Takahashi2002}
\begin{equation} \label{eq:Kex}
K_{ex}(\hat{\bm{m}}, \hat{\bm{n}})=-J_{ex}\hat{\bm{m}}\cdot \hat{\bm{n}}
\end{equation}
where $J_{ex}$ is the interface exchange coupling energy.
When $\hat{\bm{m}}=(001)$, the easy axis of the exchange coupling anisotropy energy coincides with that of the intrinsic anisotropy. Then, the total anisotropy energy becomes
\begin{equation}
\begin{split}
K_{tot}^{\hat{\bm{m}}=\hat{z}}(\theta,\mu)
=&K_i(\theta,\mu)+K_{ex}(\hat{\bm{z}}, \hat{\bm{n}}) \\
=&-K_{eff}(\mu)\sin^2\theta-J_{ex}\sin\theta. \\
\end{split}
\end{equation}
As a result, the additional exchange energy simply adds to the intrinsic anisotropy energy and N\ee{l} vector is strongly pinned to the $[001]$ direction. This results in an increase of $\mu_c$ and the AFS system remains in the gapped phase for larger range of $\mu$. 
When $\hat{\bm{m}}=(100)$, the total energy is
\begin{equation} \label{eq:Ktot}
\begin{split}
K_{tot}^{\hat{\bm{m}}=\hat{x}}(\varphi_{ex},\theta,\mu)
=&K_{i}(\theta,\mu)+K_{ex}(\hat{\bm{x}}, \hat{\bm{n}}) \\
=&-K_{eff}(\mu)\sin^2\theta-J_{ex}\cos\theta\cos\varphi_{ex},
\end{split}
\end{equation}
where $\varphi_{ex}$ is the in-plane angle between $\hat{\bm{m}}$ and $\hat{\bm{n}}$. 
The global minimum of the anisotropy energy in Eq.~(\ref{eq:Ktot}) occurs at
\begin{equation} \label{eq:phitheta}
(\varphi_{ex},\theta)=
\begin{cases}
(0,\cos^{-1}[J_{ex}/2K_{eff}]), & \text{if } 2K_{eff}>J_{ex} \\
(0,0), & \text{if } 2K_{eff}\leq J_{ex}, \\
\end{cases}
\end{equation} 
whose derivation is in Appendix~\ref{sec:GM}.
Eq.~(\ref{eq:phitheta}) describes the manner in which the N\ee{l} vector is oriented with respect to the $\hat{\bm{m}}$ for different exchange coupling energies, $J_{ex}$. When the exchange coupling energy is comparable to or smaller than the intrinsic anisotropy energy satisfying $J_{ex}<2K_{eff}$, the overall anisotropy energy is modified by the exchange coupling and gradually shifts the easy axis toward in-plane direction. Consequently, the N\ee{l} vector is tilted toward $\hat{\bm{m}}$ having an out-of-plane angle $\theta=\cos^{-1}\frac{J_{ex}}{2K_{eff}}$ for $J_{ex}<2K_{eff}$. When the exchange coupling energy is larger than $2K_{eff}$, $\hat{\bm{n}}$ is completely aligned with $\hat{\bm{m}}$. 

In our system, we assume a fixed exchange coupling energy $J_{ex}<2K_{eff}(0)$ with $\hat{\bm{m}}=(100)$. Then the N\ee{l} vector is initially tilted away from $[001]$ and the energy spectrum is gapped. As we raise the chemical potential, the magnitude of $K_{eff}(\mu)$ decreases and $\hat{\bm{n}}$ is tilted further toward $\hat{\bm{m}}$ reducing the size of the gap. When the chemical potential reaches the critical chemical potential ${\mu'}_c$ satisfying $K_{eff}(\mu_c')=J_{ex}/2$, $\hat{\bm{n}}$ is aligned with $[100]$ having the gapless spectrum. Note that such reorientation of $\hat{\bm{m}}$ occurs when $K_{eff}(\mu_c)=0$ in the absence of the exchange coupling and $|\mu_c'|<|\mu_c|$ as $K_{eff}$ is a monotonic decreasing function, as shown in Fig.~\ref{fig:FE}(a). Therefore, the topological MIT occurs at a smaller chemical potential deviation from the Dirac point when compared to the intrinsic AFS system. By contrast, when $\hat{\bm{m}}=(001)$, the spectrum remains gapped at $\mu_c$. Therefore, by comparing the energy spectrum of the two distinct $\hat{\bm{m}}$ configurations, we may observe a clear difference in the transport signature.

 \begin{figure}[t!] 
  \centering
   \includegraphics[width=0.45\textwidth]{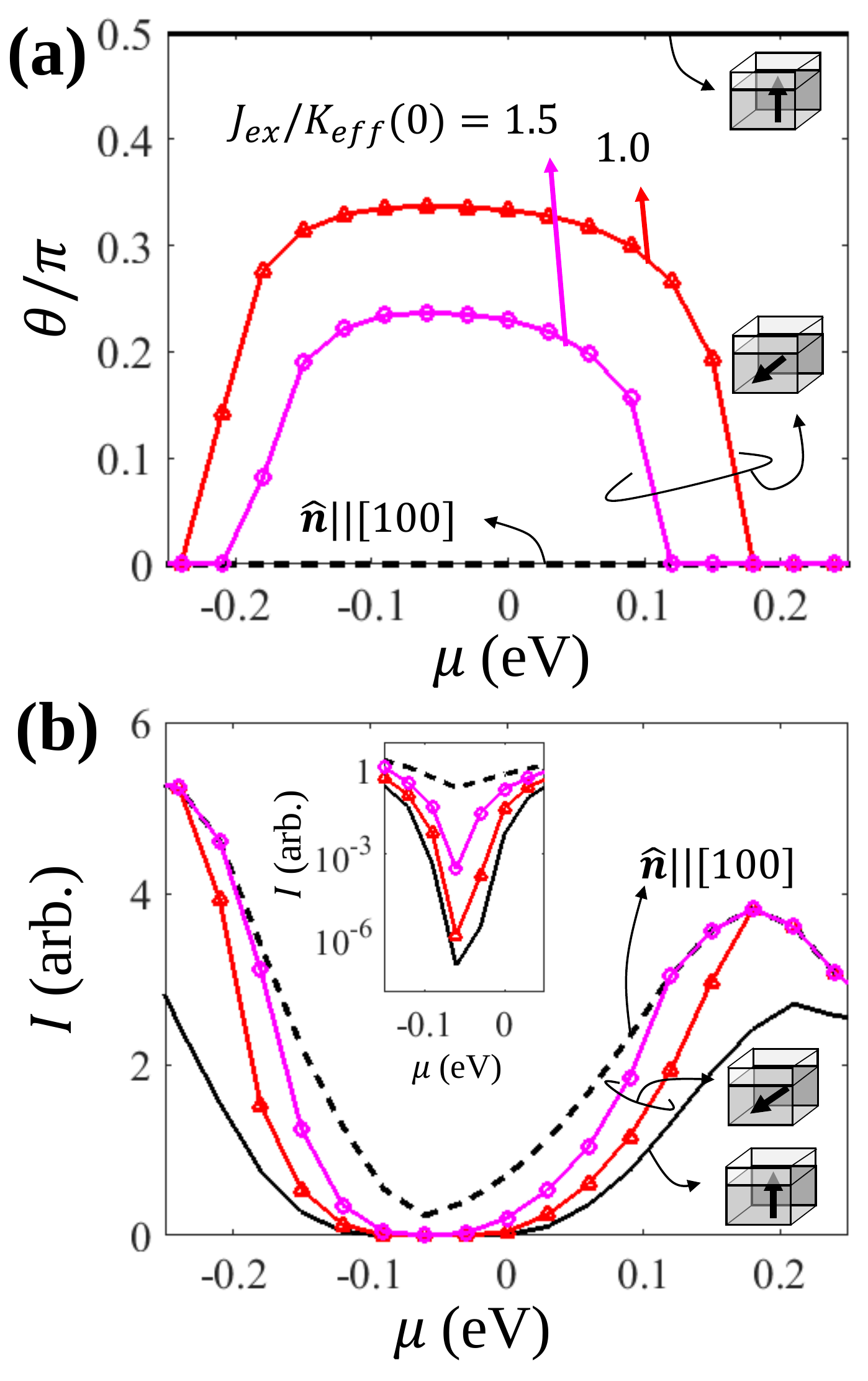}
  \caption{(a) The N\ee{l} vector orientation which minimizes the anisotropy energy in Eq.~(\ref{eq:Ktot}). The black solid line represents the AFS-FI system with $\hat{\bm{m}}||\hat{z}$. The red triangle symbol, and the magenta circular symbol represents the system with $J_{ex}=1.0K_{eff}(0)$, and $J_{ex}=1.5K_{eff}(0)$, respectively, and $\hat{\bm{m}}||\hat{x}$. (b) The current is calculated for the correspoding conditions in (a). The color and symbols are matched to describe different $J_{ex}$ strength and the current is plotted in log scale. The dashed line represents the current with $\hat{\bm{n}}||\hat{x}$ configuration. The inset shows the current near $\mu=0$ in log scale. 
  }\label{fig:I2}
\end{figure}   

To examine the transport signature of the AFS-FI system, we construct a real-space based tight-binding model. Fig.~\ref{fig:I}(a) shows a schematic of the system. We use the real-space based tight-binding model in Eq.~(\ref{eq:tb1}) and the parameters used in Fig.~\ref{fig:spectrum}(a). We set $T=77$ K in order to generate a smooth current plot, yet clearly resolve the gap in the transport results. We assume a thin slab geometry assuming that the proximity exchange coupling at the bottom surface plays a significant role throughout all layers.
The simulation geometry consists of $n_y=20$ sites in transport direction and $n_z=5$ sites for thickness direction leaving $\hat{x}$ direction in momentum space. The conductance is calculated using non-equilibrium Green's function\cite{Datta2005}. The metallic contact is connected to the left and right hand side of the device region and we use the wide-band limit (WBL) approximation\cite{Datta2005} for the contact self-energy. The interface exchange coupling is not explicitly included in our Hamiltonian. Instead, we compute the free energy of our real-space Hamiltonian from Eqs.~(\ref{eq:F}) and (\ref{eq:Phi}) and compute $K_{eff}(\mu)=-K_i(0,\frac{\pi}{2},\mu)$ as a function of $\mu$. For a given $\mu$, we assume that FI magnetization $\hat{\bm{m}}$ is aligned along the $\hat{x}$-axis with a given exchange coupling energy $J_{ex}$. Then we determine the orientation of $\hat{\bm{n}}$ from Eq.~(\ref{eq:phitheta}). 

Fig.~\ref{fig:I2}(a) shows the out-of-plane angle $\theta$ for $\hat{\bm{n}}$ as a function of $\mu$ for different values of $J_{ex}$. 
The black solid line shows $\theta$ for $\hat{\bm{m}}||[001]$. In this case, the exchange energy anisotropy is aligned with the intrinsic anisotropy energy at $\mu=0$ and $\hat{\bm{n}}$ stays pinned to $[001]$ direction. For $\hat{\bm{m}}||[100]$ case, however, Fig.~\ref{fig:I2}(a) shows an evolution of $\hat{\bm{n}}$ as a function of $\mu$. Unlike the intrinsic AFS system where $\hat{\bm{n}}$ is aligned with $\hat{z}$ at $\mu=0$, $\hat{\bm{n}}$ is tilted away from $[001]$ initially due to the non-zero $J_{ex}$. We observe that the N\ee{l} vector is further away from $[001]$ for larger exchange energy $J_{ex}$. As we increase the deviation of the chemical potential from the Dirac point, the magnitude of $K_{eff}(\mu)$ decreases and the N\ee{l} vector reorients toward in-plane direction gradually. When the reduced intrinsic anisotropy energy reaches $K_{eff}(\mu_c')=J_{ex}/2$, the N\ee{l} vector is completely aligned to $[100]$ direction and the topological MIT occurs. Such N\ee{l} vector switching occurs at smaller $\mu_c'$ for larger $J_{ex}$, consequently, Fig.~\ref{fig:I2}(a) shows that a larger interface exchange coupling energy results in a smaller critical chemical potential.

With our understanding of the interplay between intrinsic anisotropy and interface exchange coupling energy, we now examine the potential quantum transport signature. The transmission of the two terminal device is computed using $\hat{\bm{n}}$ obtained in Fig.~\ref{fig:I2}(a) for a given $\mu$ and $\hat{\bm{m}}$. Fig.~\ref{fig:I2}(b) illustrates the response of the AFS-FI structure under a small bias of $V_{LR}=10$ mV. The dashed line represents the current with $\hat{\bm{n}}||[100]$ configuration which has a gapless dispersion and thereby shows a finite current even at $\mu=0$.
The black solid line represents the current response of the AFS-FI system when $\hat{\bm{m}}||[001]$. Here, $\hat{\bm{n}}$ is pinned to $[001]$ direction and the spectrum is gapped. The corresponding current confirms that the AFS is insulating near $\mu=0$ eV.
In this case, no topological MIT occurs as $\hat{\bm{n}}$ is fixed to $[001]$ and the current is attributed to the carrier conduction through the conduction or valence band. For $\hat{\bm{m}}||[100]$ case, we observe the increased current value at $\mu=0$ (see inset of Fig~\ref{fig:I2}(b)). This is due to the fact that the gap size becomes smaller as the N\ee{l} vector is tilted toward $[100]$ as it is shown in Fig.~\ref{fig:I2}(a). For increasing $\mu$, the topological MIT occurs at $\mu_c'$ and manifests itself as a clear increase in current compared with $\hat{\bm{m}}||[001]$ case. Such distinction is more clear for larger $J_{ex}$, as the critical chemical potential $\mu_c'$ is located closer to the Dirac point where the current difference between gapped and gapless phase is maximized.


\section{Summary and Conclusion} \label{sec:Con}
We propose the voltage driven antiferromagnetic (AF) order manipulation in antiferromagnetic semimetals (AFS). We predict that the spectrum may open a gap as we reorient AF order and break the underlying non-symmorphic symmetry. By calculating the free energy of gapped and gapless spectrum, we find that the system prefers the gapped spectrum when the chemical potential is located at the Dirac point. However, the free energy difference between gapped and gapless spectrum is monotonically reduced as the chemical potential deviates from the Dirac point. We find the system eventually changes its preferred phase from gapped to gapless spectrum at the critical chemical potential. Consequently, the corresponding AF order is switched from out-of-plane to in-plane direction. Lastly, we propose a two-terminal experimental setup to identify the voltage driven N\ee{l} vector switching. The ferromagnetic insulator (FI) may be coupled to the AFS system, and introduce an additional anisotropy energy via exchange coupling in desirable direction. 
For a sufficiently large exchange energy, the N\ee{l} vector is pinned to the out-of-plane direction when the FI net magnetization is aligned along the $\hat{z}$-axis. In this case, the two terminal current shows a conduction gap that corresponds to the intrinsic gap of the gapped AFS spectrum. When the net magnetization of FI is aligned with the N\ee{l} vector of the gapless spectrum, the induced exchange energy alters the critical chemical potential. Then the N\ee{l} vector switching occurs at the reduced critical chemical potential. As a result, the two terminal current shows reduced gap-like feature which serves as an evidence for the intrinsic anisotropy energy dependence on the chemical potential of the AFS. 

The proposed AF order switching mechanism provides a simple method for electrical control over AF order, which may open a new avenue for realizing AF based spintronic device. However, one need to consider dynamic interactions between electronic structure of the AFS and N\ee{l} vector orientation as the current flow may induce additional spin-orbit torque, but we leave this for future work.

\begin{acknowledgments} 
This work is supported by the National Science Foundation (NSF) under Grant No. DMR-1720633. Y. Kim acknowledges useful initial discussions from M. J. Park. Y. Kim also thanks T. M. Philip and M. Hirsbrunner for helpful discussions.
\end{acknowledgments} 

\appendix

\section{Low-energy effective Hamiltonian near the Dirac point} \label{sec:Heff}

In this section, we obtain the low-energy effective model near the Dirac point. We begin with the Dirac point assuming that $\hat{\bm{n}}$ is aligned with $\hat{x}$ direction. Following Eq.~(\ref{eq:E2}) and discussions therein, the Dirac point is located at $\vec{k}_1=(\pi,k_{y0},k_{z0})$ where $k_{y0}$ and $k_{z0}$ satisfies $\Delta-(\lambda-\lambda_z\cos k_{z0})\sin k_{y0}=0$. We expand the Hamiltonian in Eq.~(\ref{eq:tb2}) near the momentum $\vec{k}_1$ as
\begin{equation} \label{eq:Hq}
\begin{split}
H(\vec{k}_1+\vec{q})\simeq&
(v_{x1}\tau_1+v_{x2}\tau_2+v_{x3}\tau_3\sigma_2)q_x \\
&+(v_{y}q_y+v_z q_z)\tau_3\sigma_1
+\Delta\tau_3 \bm{\sigma}\cdot \hat{\bm{n}}_1,
\end{split}
\end{equation}
where $\vec{q}=(q_x,q_y,q_z)$ is an infinitesimal deviation from $\vec{k}_1$ and $\hat{\bm{n}}_1=(\cos\theta\cos\varphi,\cos\theta\sin\varphi,\sin\theta)$ is the N\ee{l} vector deviated from $\hat{x}$ direction with the in-plane angle $\varphi$ and out-of-plane angle $\theta$. In Eq.~(\ref{eq:Hq}), 
\begin{equation} \label{eq:vxyz}
\begin{split}
v_{x1}=&-(1/2)(t_{xy}+t_z\cos k_{z0})\cos(k_{y0}/2), \\
v_{x2}=&-(1/2)t_z\sin k_{z0}\cos(k_{y0}/2), \\
v_{x3}=&-(\lambda-\lambda_z\cos k_{z0}), \\
v_{y}=&-(\lambda-\lambda_z\cos k_{z0})\cos k_{y0}, \\
v_{z}=&-\lambda_z\sin k_{z0}\sin k_{y0}. \\
\end{split}
\end{equation}
Then, the energy spectrum of Eq.~(\ref{eq:Hq}) is 
\begin{equation}
\begin{split}
E_\vec{q}=&
\left[ ( v_xq_x-\frac{v_{x3}}{v_x}\Delta_y)^2+(v_yq_y+v_zq_z-\Delta_x)^2 \right. \\
&\left. \Delta_z^2+\Delta_y^2\frac{v_{x1}^2+v_{x2}^2}{v_x^2} \right]^{1/2}, \\
\end{split}
\end{equation}
where $v_x=\sqrt{v_{x1}^2+v_{x2}^2+v_{x3}^2}$ and $\Delta\hat{\bm{n}}=(\Delta_x,\Delta_y,\Delta_z)$. As a result, the deviation of the N\ee{l} vector from $\hat\bf{n}||[100]$ configuration develops the energy gap which satisfies the following relation:
\begin{equation} \label{eq:Egap}
E_g\propto\Delta\sqrt{\sin^2\theta + \cos^2\theta\sin^2\varphi \frac{v_{x1}^2+v_{x2}^2}{v_x^2}}.
\end{equation}

\section{Free energy for fermions} \label{sec:FE}
In this section, we derive the thermodynamic potential, or the Helmholtz free energy used in this study.
The Helmholtz free energy is defined as\cite{Pathria1996}
\begin{equation}
F=-\frac{1}{\beta}\ln Z,
\end{equation}
where $\beta=1/k_B T$, $T$ is temperature, and $Z$ is the partition function. The partition function is defined as
\begin{equation}
z=e^{-\beta H}.
\end{equation}
Assuming that the Hamiltonian is diagonalized, we may write
\begin{equation}
\begin{split}
z=&e^{-\beta H}=\text{Tr}\{ e^{-\beta\sum_{\mathbf{k}}H_{\mathbf{k}}} \} \\
=&\sum_{\{n\}} e^{-\beta n_{\mathbf{k},i}\sum_{\mathbf{k},i}(E_{\mathbf{k},i}-\mu)}
=\sum_{\{n\}}\prod_{\mathbf{k},i}e^{-n_{\mathbf{k},i}\beta (E_{\mathbf{k},i}-\mu)}
\end{split}
\end{equation}
where $n_{\mathbf{k},i}=0,\;1$ is fermion occupation number at the specific state, $\mu$ is the chemical potential, and the last summation $\sum_{\{n\}}$ runs over all eigenvalues $E_{\mathbf{k},i}$, at momentum $\mathbf{k}$, which belongs to the $i$-th band with the total particle number $N$, or $\sum_{\{n\}}n_{\mathbf{k},i}=N$. 
If we consider every possible configurations, we have the grand partition function
\begin{equation}
\begin{split}
Z=&\sum_{N=0}^{\infty}\sum_{\{n\}}\prod_{\mathbf{k},i}e^{-n_{\mathbf{k},i}\beta (E_{\mathbf{k},i}-\mu)} \\
=& \prod_{\mathbf{k},i} \left[ \sum_{\{n\}}e^{-n_{\mathbf{k},i}\beta (E_{\mathbf{k},i}-\mu)}  \right]
=\prod_{\mathbf{k},i} \left( 1 + e^{-\beta (E_{\mathbf{k},i}-\mu)}  \right).
\end{split}
\end{equation}
As a result, 
\begin{equation}
F=-\frac{1}{\beta}\sum_{\mathbf{k},i}\ln\left( 1+e^{-\beta (E_{\mathbf{k},i}-\mu)} \right).
\end{equation}
In continuum limit, we use the fact that $\sum_\vec{k}=(V/(2\pi)^3)\int d^3k$. Then, 
\begin{equation} \label{eq:Fapp}
\frac{F}{V}=-\frac{1}{\beta(2\pi)^3}\int d^3 k \sum_i\ln\left( 1+e^{-\beta (E_{\mathbf{k},i}-\mu)} \right).
\end{equation}

\section{Zero-temperature limit solution for free energy} \label{sec:FElin}
The free energy function derived in Eq.~(\ref{eq:Fapp}) may be rewritten as
\begin{equation}
\frac{F}{V}=\frac{1}{(2\pi)^3}\int d^3 k \sum_i f_{\mathbf{k},i}
\end{equation}
where we define 
\begin{equation} \label{eq:fapp}
f_{\mathbf{k},i}=-\frac{1}{\beta}\ln\left( 1+e^{-\beta (E_{\mathbf{k},i}-\mu)} \right).
\end{equation}
Eq.~(\ref{eq:fapp}) represents the individual eigenstate contribution to the total free energy. In order to understand how eigenvalues change the free energy, we may linearize Eq.~(\ref{eq:fapp}) by assuming a zero-temperature limit, or $T\rightarrow 0$. Specifically, we consider the three different energy ranges shown in Fig.~\ref{fig:Fmu}(a): (i) When the eigenvalues are above the chemical potential, or $x=\beta(\mu-E_{\vec{k},i})\ll -1$, 
\begin{equation}
f_{\vec{k},i}\simeq -(1/\beta)\ln(1)=0.
\end{equation}
(ii) When the eigenvalues are near the chemical potential, or $|x|\ll 1$, 
\begin{equation}
\begin{split}
f_{\vec{k},i}\simeq& -(1/\beta)\ln\left( 2+x \right)\simeq -(1/\beta) \ln2\left( 1+\frac{1}{2\ln2}x \right) \\
\simeq& - \frac{\mu-E_{\vec{k},i}}{2},
\end{split}
\end{equation}
where we assume $\beta\rightarrow\infty$ and $(1/\beta)\ln2\rightarrow 0$.
(iii) When the eigenvalues are well below the chemical potential, or $x\gg 1$,
\begin{equation}
f_{\vec{k},i}\simeq -(1/\beta)\ln\left( e^{x} \right)= -(\mu-E_{\vec{k},i}).
\end{equation}
We summarize the results in Eq.~(\ref{eq:flin}).

\section{Finding a global minimum for anisotropy energy} \label{sec:GM}
In Section~\ref{sec:current}, we discuss about the total anisotropy energy of the AFS-FI coupled system. The total anisotropy energy in Eq.~(\ref{eq:Ktot}) reads
\begin{equation} \label{eq:Ktot2}
K_{tot}^{\hat{\bm{m}}=\hat{x}}(\varphi_{ex},\theta,\mu)
=-K_{eff}(\mu)\sin^2\theta - J_{ex}\cos\theta\cos\varphi_{ex}.
\end{equation}
For a given chemical potential $\mu$, we find $(\varphi_{ex},\theta)$ that minimizes Eq.~(\ref{eq:Ktot2}). As a first step, we find the local minima for given $\varphi_{ex}$ by varying $\theta$. Taking first derivative with respect to $\theta$, Eq.~(\ref{eq:Ktot2}) becomes
\begin{equation} \label{eq:partialK}
\begin{split}
\partial_\theta K_{tot}^{\hat{\bm{m}}=\hat{x}}(\varphi_{ex},\theta,\mu)
=&-2K_{eff}\sin\theta\cos\theta+J_{ex}\sin\theta\cos\varphi_{ex} \\
=&\sin\theta(J_{ex}\cos\varphi_{ex}-2K_{eff}\cos\theta).
\end{split}
\end{equation}
We find the local minima when $(\varphi_{ex},\theta)$ satisfies $\sin\theta=0$ or $J_{ex}\cos\varphi_{ex}-2K_{eff}\cos\theta$. When $2K_{eff}>J_{ex}$, we have the following local minima at $\theta=0$ or $\theta=\theta_1=\cos^{-1}(J_{ex}\cos\varphi_{ex}/2K_{eff})$. Each of which produces minima in anisotropy energy as
\begin{equation}
K_{tot}^{\hat{\bm{m}}=\hat{x}}=
\begin{cases}
-J_{ex}, & \text{if } \theta=0 \\
-K_{eff}\left[1+\left(\frac{J_{ex}}{2K_{eff}}\right)^2\right], &
 \text{if } \theta=\theta_1. \\
\end{cases}
\end{equation}
Among two values, global minima is found when $(\varphi_{ex},\theta)=(0,\cos^{-1}[J_{ex}/2K_{eff}])$.

When $2K_{eff}<J_{ex}$, we may find the in-plane angle $\varphi_{ex}=\varphi_{2}$ which satisfies $2K_{eff}=J_{ex}\cos\varphi_{2}$. Then, we still can find local minima which satisfies $J_{ex}\cos\varphi_{ex}-2K_{eff}\cos\theta=0$ in Eq.~(\ref{eq:partialK}).
Then we find minimum anisotropy energy as 
\begin{equation}
K_{tot}^{\hat{\bm{m}}=\hat{x}}=
\begin{cases}
-J_{ex}, & \text{if } \theta=0 \\
-2K_{eff}, &
 \text{if } \theta
 =\theta_2, \\ 
\end{cases}
\end{equation}
where $\theta_2=\cos^{-1}(J_{ex}\cos\varphi_{2}/2K_{eff})$ and $2K_{eff}=J_{ex}\cos\varphi_{2}$. Due to the condition given as $2K_{eff}<J_{ex}$, we have the global minima $K_{tot}^{\hat{\bm{m}}=\hat{x}}=-J_{ex}$ at $(\varphi_{ex},\theta)=(0,0)$. In summary, the N\ee{l} vector orientation that gives us the global minima in anisotropy energy is at
\begin{equation} \label{eq:phitheta2}
(\varphi_{ex},\theta)=
\begin{cases}
(0,\cos^{-1}[J_{ex}/2K_{eff}]), & \text{if } 2K_{eff}>J_{ex} \\
(0,0), & \text{if } 2K_{eff}\leq J_{ex}, \\
\end{cases}
\end{equation} 
which is shown in Eq.~(\ref{eq:phitheta}).


\bibliography{reference}		

\end{document}